\documentclass[aps,pra,reprint,superscriptaddress,nofootinbib,floatfix]{revtex4-2}

\usepackage{amsmath,amssymb,bm}
\usepackage{graphicx}
\usepackage[hidelinks]{hyperref}
\makeatletter
\renewcommand\section{\@startsection{section}{1}{\z@}{12pt plus 2pt minus 2pt}{6pt}{\normalfont\small\bfseries\centering}}
\makeatother
\hypersetup{pdftitle={Multipole Decomposition of Moving Sources},
pdfauthor={Shuhang Chen, Yutong Li, Xinmeng Wang, Qi Zhu}}

\begin{document}

\title{Multipole Decomposition of Moving Sources}

\author{Shuhang Chen}
\author{Yutong Li}
\author{Xinmeng Wang}
\author{Qi Zhu}
\email[Contact author: ]{zhuqi@ustc.edu.cn}
\affiliation{Department of Electronic Engineering and Information Science,
University of Science and Technology of China, Hefei, Anhui 230027, China}

\begin{abstract}
Cartesian multipole moments are widely used to analyze and engineer radiation,
but the usual relations that give the radiated field from these moments assume
a wavevector-independent source frequency. For moving sources, Doppler shifts
make the frequency depend on the wavevector, so the radiated fields predicted from the moments generally carry errors
proportional to $v/c$, where $v$ is the speed of the source. From charge
conservation in the laboratory frame, we derive a multipole decomposition of
moving sources that retains the usual moments but evaluates their
contributions at Doppler-shifted wavenumbers and adds a convection term. Requiring no rest frame, it applies to moving matter and to excitation patterns in synthetic motion
through media at rest, superluminal patterns included. The decomposition also extends to nonuniform translation and to nonrigid
patterns with curved spectral supports. Exact calculations for moving electric, magnetic, and toroidal dipoles confirm
the decomposition, which attributes the difference between the radiation of a
dipole in moving matter and in synthetic motion to the laboratory-frame moments
that each kind of motion induces. For an anapole, these moments decide whether it
radiates. An anapole carried by uniformly moving matter remains nonradiating,
whereas an excitation pattern with the same electric and toroidal dipoles
radiates in synthetic motion. The decomposition thus makes multipolar analysis available for both kinds of
motion and opens the way to designing the radiation of moving sources.
\end{abstract}

\maketitle
\raggedbottom

\section{Introduction}
\label{sec:intro}

The multipole expansion expresses the far field of a localized
charge--current distribution as a superposition of multipolar
contributions, characterized by Cartesian moments such as the
electric dipole (ED), magnetic dipole (MD), and toroidal dipole (TD) and their higher-order counterparts~\cite{GrahnShevchenkoKaivola2012,RadescuVaman2002,AlaeeRockstuhlFernandezCorbaton2019,DubovikTugushev1990,MarengoDevaney1998,LeBoudec2024}.
It is routinely used to identify the multipolar character of Mie
resonances~\cite{EvlyukhinEtAl2012,KuznetsovEtAl2016}, to characterize
scatterers and
metasurfaces~\cite{AlaeeRockstuhlFernandezCorbaton2019,TerekhovEtAl2019,KolkowskiSehrawatShevchenko2026},
and to engineer radiation through interference among multipolar fields, as
in Huygens sources and other Kerker-type
scatterers~\cite{KerkerWangGiles1983,FuEtAl2013,DeckerEtAl2015,LiuKivshar2018,ShamkhiEtAl2019},
anapoles~\cite{MiroshnichenkoEtAl2015,BaryshnikovaEtAl2019,SavinovEtAl2019CommPhys},
and bound states in the
continuum~\cite{HsuEtAl2016,KoshelevEtAl2018PRL,SadrievaEtAl2019,KoshelevEtAl2020Science}.
These applications usually concern stationary sources.  Moving
sources, however, are becoming increasingly relevant in photonics.  In
some cases, matter itself moves, as in light
sails~\cite{Atwater2018,Ilic2019,Michaeli2025} and relativistically moving
scatterers~\cite{WhittamEtAl2023,VavilinEtAl2025}.  In others, the motion
is synthetic~\cite{HarwoodEtAl2025}, with an excitation or modulation pattern
moving through stationary matter, as in space--time-modulated
metamaterials~\cite{Caloz2020a,CalozDeckLeger2020PartII,Huidobro2019,Galiffi2022,Shaltout2019a,Liberal2026,Nasari2026} and
in media driven by flying-focus pulses~\cite{Froula2018,SainteMarie2017} or
space--time wave packets~\cite{KondakciAbouraddy2017,Yessenov2022}.
Such patterns can mimic moving matter, as in Fresnel drag~\cite{Huidobro2019}
and rotation~\cite{Liberal2026,Nasari2026}.
Both kinds of motion raise the question of whether the multipole decomposition of stationary sources
carries over to moving ones.

This decomposition does not carry over unchanged, even for the simplest source, an ED.
For an ED in matter moving at constant velocity $\mathbf v$ perpendicular to
the dipole, the multipole relations of stationary sources, applied to the moments at a
given time, predict that the current components radiating forward and backward
differ by a fraction of about $v/c$, whereas they are in fact equal.
Such errors grow with the speed and become substantial for light
sails~\cite{Atwater2018} and for synthetic motion, in which the pattern can approach or even exceed
the speed of light~\cite{HarwoodEtAl2025,Froula2018}.
The same relations fail for an anapole in moving matter. The multipolar
radiation fields obtained from them do not cancel, whereas the
anapole remains nonradiating at any constant velocity~\cite{Savinov2018}.
Being proportional to $v/c$, these errors are of the same order as the
relative Doppler shift, the leading kinematic effect of the motion, and
therefore cannot be neglected.

\begin{figure}[t]
\centering
\includegraphics[width=\columnwidth]{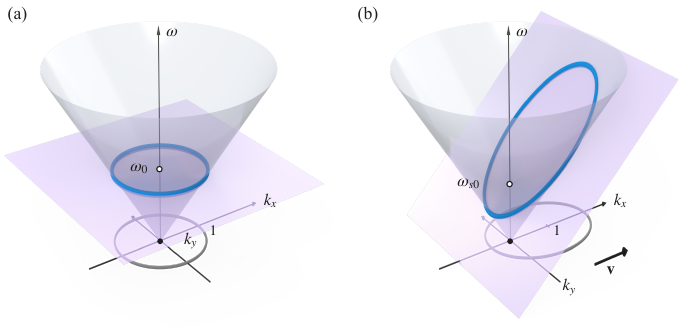}
\caption{Spectral support of a localized source (gray plane) and its
intersection with the light cone $\omega=c|\mathbf k|$ (blue curves), shown in
$(k_x,k_y,\omega)$ with wavevectors in units of $k_0$; gray curves on the base
are projections of the intersection onto the $\mathbf k$ plane.  (a) A
stationary source has the horizontal support $\omega=\omega_0$; the projection
is the circle $|\mathbf k|=k_0$, a cross section of a sphere, so the source radiates at the same wavenumber in every direction.  (b) A source moving
with $\mathbf v=c\beta\hat{\mathbf x}$, $\beta=0.6$, has the support
$\omega=\omega_{s0}+\mathbf v\cdot\mathbf k$ of slope $\mathbf v$
[Eq.~\eqref{eq:tilted-support}], with $\omega_{s0}=\omega_0/\gamma$ in moving
matter and $\omega_{s0}=\omega_0$ in synthetic motion, shown for moving matter; the projection is an ellipse displaced toward
$+\hat{\mathbf x}$, a cross section of a spheroid, and the radiation in the direction $\hat{\mathbf n}$ has the Doppler wavenumber $k(\hat{\mathbf n})$ of
Eq.~\eqref{eq:doppler-k}.}
\label{fig:support}
\end{figure}

The errors originate in applying the multipole relations of stationary sources unchanged.  These relations assume that every plane-wave component of the source
oscillates at the same frequency $\omega_0$, so that charge conservation
converts current moments into the charge moments, the ED and the electric
quadrupole (EQ), through the same factor $-i\omega_0$ for every component.  The assumption also gives every direction the same radiated wavenumber,
$k_0=\omega_0/c$.  Under this assumption, the spectral support of the source, the surface
$\omega=\omega_s(\mathbf k)$ on which its space--time spectrum
lies~\cite{SaariReivelt2004,YessenovEtAl2019,Yessenov2022},
is the horizontal plane $\omega=\omega_0$, which meets the light cone
$\omega=c|\mathbf k|$ on a sphere whose radius is the common wavenumber
$k_0$ [Fig.~\ref{fig:support}(a)].  Motion tilts this plane.  By the Doppler effect, the support becomes
$\omega=\omega_s(\mathbf 0)+\mathbf v\cdot\mathbf k$
[Fig.~\ref{fig:support}(b)].  As a result, the conversion factor varies
with the wavevector, and the radiated wavenumber varies with direction.  A multipole decomposition of moving sources must therefore account for this tilt.

Existing treatments either prescribe moving point multipoles and derive their
fields~\cite{Savinov2018,Ellis1966,Heras1998MovingDipoles,Heras1998Toroidal},
or undo the tilt by a Lorentz transformation to the rest frame of the source,
solve the scattering problem there~\cite{WhittamEtAl2023,VavilinEtAl2025}, and
transform the result back.  Both give the exact total field, but the first starts from point multipoles
chosen in advance instead of decomposing the current of an arbitrary source,
and the second yields the moments of the rest frame.  Yet identifying the
multipolar content of the radiation, or engineering it through multipolar
interference, requires a decomposition into the multipoles of the laboratory frame, where the radiation is ultimately observed.  The laboratory-frame
moments themselves follow from the Lorentz transformation, which assigns an ED to a moving MD~\cite{Fisher1971,Hnizdo2012,GratusBanaszek2018,DubovikSahaRubin2000}
but leaves the contributions of these moments to the radiation undetermined.
Determining these contributions with the multipole relations of stationary
sources produces the errors described above.  Moreover, a pattern moving faster than
light has no rest frame to transform to, and an accelerated source has a
different one at each instant.  For both reasons, moving sources call for such a decomposition, formulated
directly in the laboratory frame.  There, however, the
current at one instant lacks the time dependence imposed by the
motion~\cite{Savinov2018} and so does not fix the frequency of each plane-wave
component.  Higher moments about a fixed origin grow as the source moves away from it,
so that the expansion about that origin cannot be truncated~\cite{VavilinEtAl2025}.  A frequency-domain multipole decomposition in the laboratory frame that can
be truncated at low order, the form in which multipolar radiation is identified and engineered,
has therefore been available only for stationary sources.

Here we establish the multipole decomposition of moving sources directly in
the laboratory frame, building it on the tilted support instead of undoing the
tilt.  The tilt itself supplies the time dependence that the current at one
instant lacks, assigning each plane-wave component its Doppler-shifted frequency.  Measured
from the instantaneous center, the moments of a translating source change in time only by a carrier phase, so those at one instant determine the radiation.  The
same tilt changes the multipole relations in two ways.  The varying conversion
factor adds to the EQ term a convection term arising from the current carried by the moving ED, and
each multipole radiates at the direction-dependent Doppler wavenumber.  Both
changes involve only laboratory-frame quantities, so the decomposition serves
moving matter and synthetic motion alike, superluminal patterns included, and
extends to all orders and to nonuniform and nonrigid motion.  Tested on an ED, an MD, and a TD in both kinds of motion, it
reproduces the exact radiation.  Carried by moving matter, an anapole acquires
an MD and an EQ yet remains nonradiating, because the two changes complete the
cancellation of its multipolar fields.  In synthetic motion, an excitation
pattern with the same ED and TD acquires a different MD and no EQ, so the
cancellation breaks and the pattern radiates.
Synthetic motion thus does not mimic moving matter in the radiation of a localized source with the same ED and TD.
These results bring multipolar analysis to sources in which matter or a
pattern moves.

\section{Multipole relations for a moving source}
\label{sec:relations}

In the laboratory frame, the plane-wave components of a moving source
oscillate at different frequencies, so we describe the source component by
component.  A localized source translating at constant velocity
$\mathbf v=c\boldsymbol\beta$, where $c$ is the speed of light, is a distribution that moves with $\mathbf v$ while oscillating at a carrier
frequency $\omega_{s0}$. Expressing the distribution at $t=0$ through its
spatial Fourier transforms $\widetilde\rho(\mathbf k)$ and $\widetilde{\mathbf J}(\mathbf k)$,
we obtain the charge density at position $\mathbf r$ and time $t$,
\begin{equation}
 \begin{aligned}
 \rho(\mathbf r,t)&=\rho(\mathbf r-\mathbf vt,0)\,e^{-i\omega_{s0}t}\\
 &=\int\!\frac{d^3k}{(2\pi)^3}\,\widetilde\rho(\mathbf k)\,
 e^{i[\mathbf k\cdot(\mathbf r-\mathbf vt)-\omega_{s0}t]}\\
 &=\int\!\frac{d^3k}{(2\pi)^3}\,\widetilde\rho(\mathbf k)\,
 e^{i[\mathbf k\cdot\mathbf r-\omega_s(\mathbf k)t]},
 \end{aligned}
 \label{eq:translating-source}
\end{equation}
and likewise the current density $\mathbf J(\mathbf r,t)$ with
$\widetilde{\mathbf J}(\mathbf k)$.  The component with wavevector
$\mathbf k$ oscillates at
\begin{equation}
 \omega_s(\mathbf k)=\omega_{s0}+\mathbf v\cdot\mathbf k.
 \label{eq:tilted-support}
\end{equation}
Moving matter and synthetic motion share this form and differ in the carrier
frequency and in whether the matter moves with the pattern.  For moving matter, in which the
charges that form the source move with $\mathbf v$, the carrier frequency is lowered by time dilation to $\omega_{s0}=\omega_0/\gamma$, where $\omega_0$
is the comoving-frame frequency of the source,
$\gamma=(1-\beta^2)^{-1/2}$, and $\beta=|\boldsymbol\beta|$.  For synthetic
motion, the excitation pattern moves through matter at rest, so
$\omega_{s0}=\omega_0$ is the excitation frequency and $\beta$ may exceed 1.
Here the pattern is described by the laboratory polarization and magnetization
it produces, including the response of the matter (Appendix~\ref{app:sources}).
In both cases the spectrum lies on the tilted plane $\omega=\omega_s(\mathbf k)$
of Eq.~\eqref{eq:tilted-support} [Fig.~\ref{fig:support}(b)].

Not every component on this plane radiates.  A plane wave propagates in vacuum
only if its frequency obeys the dispersion relation $\omega=c|\mathbf k|$, so a
component radiates only if it lies on the light cone.  For the direction given
by a unit vector $\hat{\mathbf n}$ we set $\mathbf k=k\hat{\mathbf n}$, and
Eq.~\eqref{eq:tilted-support} turns the dispersion relation into
$ck=\omega_{s0}+k\,\mathbf v\cdot\hat{\mathbf n}$, whose solution, the Doppler
wavenumber, is
\begin{equation}
 k(\hat{\mathbf n})=\frac{\omega_{s0}}{c(1-\boldsymbol\beta\cdot\hat{\mathbf n})}
 \label{eq:doppler-k}
\end{equation}
for $\boldsymbol\beta\cdot\hat{\mathbf n}<1$, which covers every direction when
$\beta<1$; directions with $\boldsymbol\beta\cdot\hat{\mathbf n}>1$ occur only
for patterns faster than light and are treated in Sec.~\ref{sec:scope}.  For
moving matter $k(\hat{\mathbf n})=k_0/[\gamma(1-\boldsymbol\beta\cdot\hat{\mathbf n})]$,
and for synthetic motion
$k(\hat{\mathbf n})=k_0/(1-\boldsymbol\beta\cdot\hat{\mathbf n})$, with
$k_0=\omega_0/c$.  As $\hat{\mathbf n}$ varies, $k(\hat{\mathbf n})\hat{\mathbf n}$
traces a sphere of radius $k_0$ for a stationary source and a spheroid displaced
toward the direction of motion for a moving source with $\beta<1$.
Figure~\ref{fig:support} shows their cross sections, a circle and an ellipse,
with one spatial dimension suppressed.  At the wavevector
$k(\hat{\mathbf n})\hat{\mathbf n}$, only the part of the current transverse to
$\hat{\mathbf n}$ radiates~\cite{Jackson1999,DevaneyWolf1973} (Appendix~\ref{app:power}).  This part is
$\Pi(\hat{\mathbf n})\widetilde{\mathbf J}(k(\hat{\mathbf n})\hat{\mathbf n})$,
where $\Pi(\hat{\mathbf n})=I-\hat{\mathbf n}\hat{\mathbf n}^{\mathsf T}$ and
$I$ is the identity.  A multipole decomposition of the radiation in the direction $\hat{\mathbf n}$
must therefore split this transverse current, taken at a wavenumber that varies
with the direction $\hat{\mathbf n}$, into multipole terms, whereas for a
stationary source the wavenumber is $k_0$ in every direction.

To bring in the multipole moments, we expand the source amplitudes $\widetilde\rho$ and $\widetilde{\mathbf J}$
about $\mathbf k=0$. The expansion is accurate when the source is much smaller
than the radiated wavelength $2\pi/k(\hat{\mathbf n})$.  Expanding $e^{-i\mathbf k\cdot\mathbf r}$ in the
Fourier transforms gives, as in the standard multipole
expansion~\cite{Jackson1999},
\begin{equation}
 \begin{aligned}
 \widetilde\rho(\mathbf k)&=q-i\,\mathbf p\cdot\mathbf k-\tfrac12S_{ab}k_ak_b+\cdots,\\
 \widetilde J_a(\mathbf k)&=\widetilde J_a(0)-iM_{ab}k_b+\cdots,
 \end{aligned}
 \label{eq:taylor-moments}
\end{equation}
where subscripts $a,b\in\{1,2,3\}$ label Cartesian components and repeated
subscripts are summed.  The coefficients are the total charge $q=\int\rho\,d^3r$, the
ED $\mathbf p=\int\mathbf r\rho\,d^3r$, the second charge moment
$\mathsf S$ with components $S_{ab}=\int r_ar_b\rho\,d^3r$, and the first
current moment with components $M_{ab}=\int r_bJ_a\,d^3r$, all for the source
at $t=0$.  Because the expansion involves only the spatial distribution of
the source at one instant, and no frequency, these moments are defined and
computed exactly as for a stationary source.  They are evaluated in the
laboratory frame, with the position vector $\mathbf r$ measured from the
instantaneous center of the source; this shifts the origin but does not change
the frame.  With this choice and a constant velocity, the moments at time $t$
differ from those at $t=0$ only by the factor $e^{-i\omega_{s0}t}$, so the
moments at $t=0$ determine those at all times.  About a fixed origin, the higher moments of a translating source would grow
with its distance from the origin, and the expansion could not be
truncated~\cite{VavilinEtAl2025}.

The multipole relations of stationary
sources~\cite{RadescuVaman2002,AlaeeRockstuhlFernandezCorbaton2019} express
this current through the ED, MD, TD, and EQ, of which the ED and EQ are defined
from the charge distribution.
The current coefficients $\widetilde{\mathbf J}(0)$ and $M_{ab}$ in
Eq.~\eqref{eq:taylor-moments} must therefore be rewritten in terms of these
moments.  We split the first current moment into an antisymmetric part,
$\tfrac12(M_{ab}-M_{ba})=\tfrac12\int(r_bJ_a-r_aJ_b)\,d^3r$, and a symmetric
part, $\tfrac12(M_{ab}+M_{ba})=\tfrac12\int(r_bJ_a+r_aJ_b)\,d^3r$.  The
antisymmetric part carries only the information in $\mathbf r\times\mathbf J$:
with $(\mathbf k\cdot\mathbf r)\mathbf J-(\mathbf k\cdot\mathbf J)\mathbf r
=-\mathbf k\times(\mathbf r\times\mathbf J)$, its contribution to
Eq.~\eqref{eq:taylor-moments} is $ic\,\mathbf k\times\mathbf m$, the radiating
term of the MD
$\mathbf m=(2c)^{-1}\int\mathbf r\times\mathbf J\,d^3r$ in Gaussian
units~\cite{Jackson1999}.  At the next order in $\mathbf k$, the current
expansion yields the TD
$\mathbf T=(10c)^{-1}\int[\mathbf r(\mathbf r\cdot\mathbf J)-2r^2\mathbf J]\,d^3r$,
with $r=|\mathbf r|$~\cite{RadescuVaman2002,AlaeeRockstuhlFernandezCorbaton2019}.
The symmetric part and $\widetilde{\mathbf J}(0)$ do not reduce to
$\mathbf r\times\mathbf J$.  They are converted into charge moments through
charge conservation, $\partial_t\rho+\nabla\cdot\mathbf J=0$, which ties the
charge density to the longitudinal part of the current (Appendix~\ref{app:power}).  Because each component oscillates at its own frequency, the
conversion factor depends on the wavevector; on the spectral support, charge
conservation reads
\begin{equation}
 \mathbf k\cdot\widetilde{\mathbf J}(\mathbf k)
 =\omega_s(\mathbf k)\,\widetilde\rho(\mathbf k).
 \label{eq:fourier-continuity}
\end{equation}
Substituting Eq.~\eqref{eq:taylor-moments}, the left-hand side becomes
$k_a\widetilde J_a(0)-iM_{ab}k_ak_b+\cdots$, and the right-hand side, the
product of $\omega_s(\mathbf k)=\omega_{s0}+v_ak_a$ with $\widetilde\rho(\mathbf k)$,
becomes
\begin{equation}
 \omega_{s0}q+\big(qv_a-i\omega_{s0}p_a\big)k_a
 +\big(-\tfrac12\omega_{s0}S_{ab}-iv_ap_b\big)k_ak_b+\cdots.
 \label{eq:continuity-expanded}
\end{equation}
The term $-iv_ap_b$ is the cross term between $\mathbf v\cdot\mathbf k$ and
$-i\,\mathbf p\cdot\mathbf k$; it is absent for a stationary source.  Matching
the terms of zeroth, first, and second order in $\mathbf k$, where only the
symmetric part of each second-order coefficient contributes, gives
\begin{equation}
 \begin{gathered}
 \omega_{s0}q=0,\qquad
 \widetilde{\mathbf J}(0)=-i\omega_{s0}\mathbf p,\\
 M_{ab}+M_{ba}=-i\omega_{s0}S_{ab}+v_ap_b+v_bp_a.
 \end{gathered}
 \label{eq:moment-identity}
\end{equation}
The first relation shows that a source with a nonzero carrier frequency is
neutral, i.e., $q=0$, which removes the term $qv_a$ from the first-degree
matching; the second then states that its net current is that of its
oscillating ED.  In the third, the term $-i\omega_{s0}S_{ab}$ is
the conversion already present for a stationary source, which turns the
symmetric current moment into the traceless EQ
$\mathsf Q=3\mathsf S-(\operatorname{tr}\mathsf S)I$.  The remaining terms,
$v_ap_b+v_bp_a$, appear only because the support is tilted: they arise from
its slope $\mathbf v$ [Fig.~\ref{fig:support}(b)] and represent a convection current, the current carried by the ED as the source
moves~\cite{Ribaric1995}.

Substituting Eq.~\eqref{eq:moment-identity} into the current expansion of
Eq.~\eqref{eq:taylor-moments} and projecting with $\Pi=\Pi(\hat{\mathbf k})$,
transverse to $\hat{\mathbf k}=\mathbf k/|\mathbf k|$, gives the multipole
expansion of the transverse current of a moving source (Appendix~\ref{app:continuity}).  The ED term comes from $\widetilde{\mathbf J}(0)=-i\omega_{s0}\mathbf p$,
and the MD term $ic\,\mathbf k\times\mathbf m$ from the
antisymmetric part of $M_{ab}$.  In the symmetric part, $-i\omega_{s0}S_{ab}$
contributes $-\tfrac12\omega_{s0}\Pi(\mathsf S\mathbf k)$; the trace part of
$\mathsf S$ contributes a vector parallel to $\mathbf k$, which the projection
removes, leaving $-\tfrac16\omega_{s0}\Pi(\mathsf Q\mathbf k)$ (Appendix~\ref{app:continuity}).  The convection current contributes
$-\tfrac{i}{2}\Pi[\mathbf v(\mathbf p\cdot\mathbf k)+\mathbf p(\mathbf v\cdot\mathbf k)]$,
and the second current moment, which enters Eq.~\eqref{eq:taylor-moments} at
second order in $\mathbf k$, contributes the TD term. The magnetic quadrupole
(MQ) and higher multipoles follow likewise (Appendix~\ref{app:continuity}).  Writing out the ED, MD, EQ, and TD terms, we obtain
\begin{align}
 \Pi\widetilde{\mathbf J}(\mathbf k)={}&
 -i\omega_{s0}\Pi\mathbf p+c|\mathbf k|^2\Pi\mathbf T
 +ic\,\mathbf k\times\mathbf m
 -\frac{\omega_{s0}}{6}\Pi(\mathsf Q\mathbf k)\notag\\
 &-\frac{i}{2}\Pi[\mathbf v(\mathbf p\cdot\mathbf k)
 +\mathbf p(\mathbf v\cdot\mathbf k)]+\cdots.
        \label{eq:moving-multipole}
\end{align}

The first four terms have
the form they take for a stationary source: for $\mathbf v=0$ and
$\mathbf k=k_0\hat{\mathbf n}$, Eq.~\eqref{eq:moving-multipole} reduces to the
ED, TD, MD, and EQ relations~\cite{RadescuVaman2002,AlaeeRockstuhlFernandezCorbaton2019}.
We refer to these relations, applied unchanged to a moving source with its laboratory-frame moments and
the fixed carrier wavenumber $\omega_{s0}/c$, as the
conventional relations.  Equation~\eqref{eq:moving-multipole} differs from the conventional relations in two
respects. It is evaluated at the wavevector
$\mathbf k=k(\hat{\mathbf n})\hat{\mathbf n}$, whose wavenumber is the Doppler
wavenumber, and it contains the term proportional to $\mathbf v$, which we
call the convection term.
Neither introduces a new multipole moment.  In Eq.~\eqref{eq:moving-multipole}, the convection term involves only the ED.
Convection terms carried by the higher charge moments, beginning with
$\mathsf S$, appear at higher order in $\mathbf k$ (Appendix~\ref{app:continuity})
and, for the sources considered here, either vanish or are of order $\beta^2$.

Both differences originate in the tilt of the support
[Fig.~\ref{fig:support}(b)]: its intersection with the light cone gives the
Doppler wavenumber, and its slope gives the convection term.  Because
Eq.~\eqref{eq:moving-multipole} is built from the laboratory-frame current
alone and requires no rest frame, it applies equally to moving matter and to
synthetic motion; the two differ only through the carrier frequency
$\omega_{s0}$ and through the laboratory-frame moments induced by the motion, which
depend on whether matter or only a pattern moves.

\begin{figure}[t]
 \centering
 \includegraphics[width=\columnwidth]{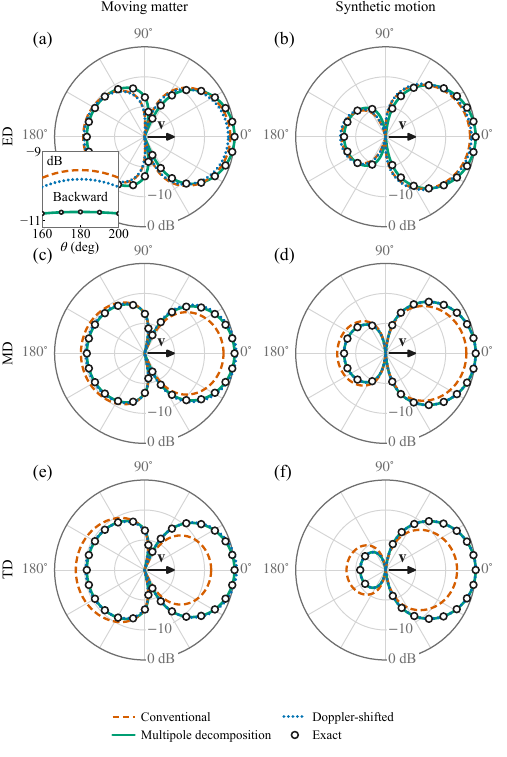}
 \caption{Radiation patterns of electric (a, b), magnetic (c, d), and toroidal
(e, f) dipoles perpendicular to $\mathbf v$ at $\beta=0.3$, in moving matter
(a, c, e) and in synthetic motion (b, d, f).  Dashed curves are obtained from the conventional relations, dotted curves
from the Doppler-shifted relations, which differ from the conventional
relations only by the evaluation at the Doppler wavenumber $k(\hat{\mathbf n})$,
solid curves from the multipole decomposition (for the TD in moving matter,
completed through the second current moment, Appendix~\ref{app:continuity}),
and circles show the exact pattern.  The plane of
each plot contains $\mathbf v$ and the dipole, and the polar angle is measured
from $\mathbf v$.  Each pattern is normalized to the maximum of the exact
pattern in its panel and shown in decibels; the inset in (a) magnifies the
backward lobe.}
 \label{fig:dipoles}
\end{figure}

\section{Radiation of elementary dipoles in moving matter and synthetic motion}
\label{sec:dipoles}

Motion changes the multipole content of a source.  In moving matter, an ED in
the comoving frame acquires an MD in the laboratory frame, an MD acquires an
ED, and a TD acquires both an MD and an EQ (Appendix~\ref{app:dipoles}).  With
Eq.~\eqref{eq:moving-multipole}, the radiation of a moving source can be
decomposed into the contributions of these laboratory-frame multipoles.  We
apply this decomposition to an ED, an MD, and a TD in both kinds of motion, and
attribute the differences between the two kinds of motion to the
laboratory-frame moments that each induces.

To check the decomposition, we compare the radiation pattern it predicts with
the exact pattern, computed directly from the laboratory-frame current without
a multipole expansion (Appendix~\ref{app:dipoles}).
Figure~\ref{fig:dipoles} shows the comparison for dipoles perpendicular to
$\mathbf v$: dashed curves are obtained from the conventional relations, dotted curves
from the Doppler-shifted relations, which differ from the conventional
relations only by the evaluation at the Doppler wavenumber, solid curves from
the multipole decomposition, and circles show the exact pattern.
The pattern is the squared magnitude of the transverse current
$\Pi\widetilde{\mathbf J}(k(\hat{\mathbf n})\hat{\mathbf n})$, the source amplitude that determines the radiation into the direction $\hat{\mathbf n}$
(Sec.~\ref{sec:relations}), multiplied by a direction-dependent factor,
identical for all curves of a given source, that carries the forward beaming
of moving sources~\cite{Jackson1999} (Appendix~\ref{app:power}).  Any
difference between a predicted and an exact pattern therefore reflects a
difference in the transverse current, whose values for
$\hat{\mathbf n}=\pm\hat{\mathbf v}$, the forward and backward transverse
currents, we use below.

We begin with the ED.  The evaluation at the Doppler wavenumber and the
convection term are both of order $\beta$, so for moving matter we work
through $O(\beta)$.  In moving matter, an ED $\mathbf d$ carries the
laboratory-frame MD $\mathbf m=-\tfrac12\boldsymbol\beta\times\mathbf d$, and
the conventional relations assign to this MD a forward--backward asymmetry:
for $\mathbf d\perp\mathbf v$, they give forward and backward transverse
currents $-i\omega_0\mathbf d(1\mp\beta/2)$.  The exact calculation, however,
gives equal forward and backward transverse currents (Appendix~\ref{app:dipoles}).  The conventional relations thus overstate the backward current
relative to the forward one, and the dashed curve in
Fig.~\ref{fig:dipoles}(a) has too large a backward lobe.  The exact pattern,
whose forward and backward currents are equal, is stronger forward only
because of the beaming factor.

Of the two changes that Eq.~\eqref{eq:moving-multipole} makes to the
conventional relations, the evaluation at the Doppler wavenumber does not
remove the asymmetry.  The ED term does not depend on $\mathbf k$, and the MD
term is already of order $\beta$, so evaluating both at $k(\hat{\mathbf n})$
changes the transverse current only at $O(\beta^2)$.  The Doppler-shifted
relations therefore differ from the conventional ones only at this order
(dotted and dashed curves).  Instead, the convection term removes the
asymmetry.
The multipole decomposition, Eq.~\eqref{eq:moving-multipole}, gives
\begin{equation}
 \begin{aligned}
 \Pi\widetilde{\mathbf J}={}&-i\omega_0\mathbf d_\perp
 -\frac{i\omega_0}{2}\big(d_n\boldsymbol\beta_\perp-\beta_n\mathbf d_\perp\big)
 \\
 &-\frac{i\omega_0}{2}\big(d_n\boldsymbol\beta_\perp+\beta_n\mathbf d_\perp\big)
 +O(\beta^2),
 \end{aligned}
 \label{eq:material-ed}
\end{equation}
where the three terms arise from the ED, the MD, and the convection term, and
$a_n=\mathbf a\cdot\hat{\mathbf n}$ and
$\mathbf a_\perp=\Pi(\hat{\mathbf n})\mathbf a$ for any vector $\mathbf a$.
The MD term contains the part $\beta_n\mathbf d_\perp$, which changes sign
between the forward and backward directions and thus causes the asymmetry.
The convection term cancels this part, so the forward and backward transverse
currents both equal $-i\omega_0\mathbf d$, as for the ED at rest.  The
multipole decomposition therefore reproduces the exact pattern [solid curve in
Fig.~\ref{fig:dipoles}(a)].  The conventional relations contain the correct MD but lack the convection term.

The MD acquired by a moving ED depends on what moves.  In moving matter, two
currents contribute to it.  The polarization current of the translating dipole contributes $+\tfrac12\boldsymbol\beta\times\mathbf d$.  Because the
charges that form the dipole are themselves carried at $\mathbf v$, the
current also contains a magnetization current, which contributes
$-\boldsymbol\beta\times\mathbf d$, the magnetic moment that a Lorentz
transformation assigns to a moving ED~\cite{Hnizdo2012}.  Together they give
$\mathbf m=-\tfrac12\boldsymbol\beta\times\mathbf d$.  In synthetic motion,
the matter stays at rest while the polarization pattern moves, so no
magnetization current arises, and the MD is
$+\tfrac12\boldsymbol\beta\times\mathbf d$, opposite in sign to that in
moving matter.  The conventional relations then give forward and backward
transverse currents $-i\omega_0\mathbf d(1\pm\beta/2)$, whereas the exact ones
are $-i\omega_0\mathbf d(1\pm\beta)$ through $O(\beta)$.  The convection term
supplies the missing half [Fig.~\ref{fig:dipoles}(b)].  In moving matter the
same term cancels the asymmetry of the MD, whereas here it doubles it.  To all
orders in $\beta$, the multipole decomposition gives the exact transverse
current $-ick(\hat{\mathbf n})\mathbf d_\perp$, whose forward-to-backward
ratio is $(1+\beta)/(1-\beta)$.  The same ED at the same speed therefore
radiates more strongly forward in synthetic motion than in moving matter
[Figs.~\ref{fig:dipoles}(a) and (b)].  Through $O(\beta)$, the ED term and the convection term are the same in both
kinds of motion, so this difference comes entirely from the MD, that is, from whether matter or a pattern moves.

We now turn to an MD and a TD, and denote by $\boldsymbol\mu$ and
$\boldsymbol\tau$ their moments in the absence of motion, to distinguish them from the laboratory-frame moments that motion induces.  For these dipoles, the
conventional relations fail for another reason, the wavenumber at which the
dipoles radiate.  The convection term, which involves only the ED, enters only
at $O(\beta^2)$, since these dipoles have no ED at rest.  The MD term grows as
$k$ and the TD term as $k^2$, so these terms are larger forward, where
$k(\hat{\mathbf n})$ exceeds $k_0$, and smaller backward.  The conventional
relations miss this variation, and the dashed curves in
Figs.~\ref{fig:dipoles}(c), (d), (e), and (f) depart from the circles.
The Doppler-shifted relations include it and, without the convection term,
already agree with the exact patterns through $O(\beta)$ (dotted curves).  The multipole decomposition coincides with the Doppler-shifted relations to
this order (solid curves).  For
the TD in moving matter, the terms written out in Eq.~\eqref{eq:moving-multipole}
leave a difference of order $\beta^2$, because the second current moment of that TD also contains a magnetic
quadrupole and a convection term carried by the EQ. With these contributions,
the multipole decomposition reproduces the exact pattern in this case as well
[Fig.~\ref{fig:dipoles}(e)] (Appendices~\ref{app:continuity} and \ref{app:dipoles}).

In synthetic motion, these dipoles acquire no other moments, so the transverse
currents $ick(\hat{\mathbf n})\,\hat{\mathbf n}\times\boldsymbol\mu$ and
$ck(\hat{\mathbf n})^2\boldsymbol\tau_\perp$ given by the multipole
decomposition are exact for any $\beta$.  The TD current grows as
$k(\hat{\mathbf n})^2$, so its forward-to-backward ratio for
$\boldsymbol\tau\perp\mathbf v$ is $[(1+\beta)/(1-\beta)]^2$, the square of
that of the ED and MD.  Its pattern is accordingly the most strongly forward
[Figs.~\ref{fig:dipoles}(b), (d), and (f)].

To leading order in $\beta$, the relations for the ED thus need only the
convection term, and those for the MD and TD need only the Doppler wavenumber.  Each deviation of the
conventional relations in Fig.~\ref{fig:dipoles} traces back to a single
missing change, and adding that change alone restores agreement with the exact pattern to this order.  In this way, the three elementary dipoles isolate the effect of each change
and show that both are necessary.

\begin{figure}[t]
 \centering
 \includegraphics[width=\columnwidth]{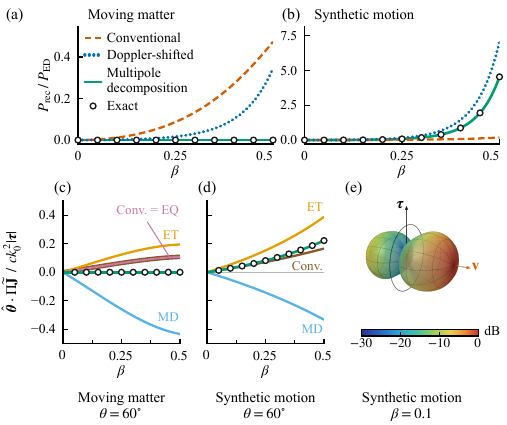}
  \caption{Moving anapole with $\boldsymbol\tau\perp\mathbf v$.  (a) Moving matter
and (b) synthetic motion: total received power versus $\beta$, in units of the
power $P_{\rm ED}=\omega_0^4|\mathbf d|^2/(3c^3)$ that its ED alone radiates at
rest; ``Conventional'' and ``Doppler-shifted'' denote the conventional and
the Doppler-shifted relations, as in Fig.~\ref{fig:dipoles}; solid curves are obtained from the multipole decomposition, completed in
moving matter through the second current moment, and circles show the exact power. Both axes are linear.
In moving matter the exact power and the power given by the multipole
decomposition vanish for all $\beta<1$.  (c),~(d) Transverse current projected on the polar unit vector
$\hat{\boldsymbol\theta}$, $\hat{\boldsymbol\theta}\cdot\Pi\widetilde{\mathbf J}$,
along $\theta=60^\circ$ in the plane of $\mathbf v$ and $\boldsymbol\tau$, in units
of $ck_0^2|\boldsymbol\tau|$, versus $\beta$, for moving matter (c) and synthetic
motion (d): the terms written out in Eq.~\eqref{eq:moving-multipole} in color
(ET denotes the ED and TD terms together, and Conv. the convection term; EQ
and Conv. coincide in moving matter), the multipole decomposition (solid), which in (c) also includes the rest of
the second current moment, and the exact
current (circles).  (e) Exact radiation pattern in synthetic
motion at $\beta=0.1$, with radius and color in decibels relative to its maximum
(30~dB range); the ring marks the plane perpendicular to $\mathbf v$.}
 \label{fig:anapole}
\end{figure}

\section{Moving anapole}
\label{sec:anapole}

An anapole combines an ED and a TD.  In Sec.~\ref{sec:dipoles}, the relations for the ED need only the convection
term and those for the TD only the Doppler wavenumber, so the
anapole is the simplest source in which both changes act together.  Its ED
$\mathbf d$ and TD $\boldsymbol\tau$ satisfy $\mathbf d=-ik_0\boldsymbol\tau$,
so that their terms in Eq.~\eqref{eq:moving-multipole},
$-i\omega_0\Pi\mathbf d$ and $ck_0^2\Pi\boldsymbol\tau$, cancel and the
anapole does not radiate at rest~\cite{MiroshnichenkoEtAl2015,BaryshnikovaEtAl2019}.
This cancellation holds only at the wavenumber $k_0$, so motion, which makes
the radiated wavenumber depend on direction, does not guarantee that a moving anapole remains nonradiating.  In this section, we determine from the
laboratory-frame multipoles whether a moving anapole radiates, both in moving
matter and in synthetic motion, for $\boldsymbol\tau\perp\mathbf v$
(Fig.~\ref{fig:anapole}).

We begin with moving matter, where the answer is already known.  The anapole remains nonradiating at any constant velocity~\cite{Savinov2018}, because a radiation field that vanishes in the rest frame vanishes in every frame~\cite{Jackson1999}.  This argument shows that the radiation vanishes, but not how the
laboratory-frame multipoles cancel.  As shown in Sec.~\ref{sec:dipoles}, the motion induces an MD and an EQ, and
Eq.~\eqref{eq:moving-multipole} contains in addition the convection term.  With
$\mathbf d=-ik_0\boldsymbol\tau$ and
$k(\hat{\mathbf n})=k_0(1+\beta_n)+O(\beta^2)$, where
$\beta_n=\boldsymbol\beta\cdot\hat{\mathbf n}$, from
Eq.~\eqref{eq:doppler-k} (Appendix~\ref{app:support}), the multipole
decomposition gives
\begin{equation}
 \begin{aligned}
 \Pi\widetilde{\mathbf J}={}&\underbrace{-ck_0^2\boldsymbol\tau_\perp}_{\text{ED}}
 +\underbrace{ck_0^2(1+2\beta_n)\boldsymbol\tau_\perp}_{\text{TD}}\\
 &+\underbrace{ck_0^2(\tau_n\boldsymbol\beta_\perp-\beta_n\boldsymbol\tau_\perp)}_{\text{MD}}\\
 &\underbrace{-\tfrac12ck_0^2(\beta_n\boldsymbol\tau_\perp+\tau_n\boldsymbol\beta_\perp)}_{\text{EQ}}\\
 &\underbrace{-\tfrac12ck_0^2(\beta_n\boldsymbol\tau_\perp+\tau_n\boldsymbol\beta_\perp)}_{\text{conv.}}
 +O(\beta^2).
 \end{aligned}
 \label{eq:anapole-terms}
\end{equation}
The ED and TD terms cancel at zeroth order, as at rest, but the TD term,
evaluated at $k(\hat{\mathbf n})$, leaves $2ck_0^2\beta_n\boldsymbol\tau_\perp$.
None of the other three terms vanishes on its own [Fig.~\ref{fig:anapole}(c)].
Yet, to this order, they sum to $-2ck_0^2\beta_n\boldsymbol\tau_\perp$ and
cancel the remainder of the TD term.

Figure~\ref{fig:anapole}(a) compares the power predicted by the conventional
relations, the Doppler-shifted relations, and the multipole decomposition with
the exact zero.  The conventional relations evaluate the TD term at
the fixed carrier wavenumber, so the ED and TD terms still cancel, but the motion-induced MD and EQ
terms remain.  The conventional relations therefore predict that the
uniformly moving anapole radiates (dashed curve).  The Doppler-shifted
relations evaluate the TD term at $k(\hat{\mathbf n})$ and so contain the ED,
TD, MD, and EQ terms, but they lack the convection term.  This term equals the
EQ term, and their curves coincide in Fig.~\ref{fig:anapole}(c).  Without it,
the Doppler-shifted relations also predict radiation (dotted curve in
Fig.~\ref{fig:anapole}(a)).  With all five terms, the terms of order $\beta$
cancel, and a remainder of order $\beta^2$ is left.  This remainder is
canceled exactly by the rest of the second current moment, because the
laboratory current of a point anapole is a polynomial of degree two in
$\mathbf k$ (Appendix~\ref{app:continuity}).  The multipole decomposition then
gives no radiation for any $\beta<1$, as the exact calculation does [solid
curves and circles in Figs.~\ref{fig:anapole}(a) and \ref{fig:anapole}(c)].
The laboratory-frame current at a single instant thus accounts for the absence of radiation from the uniformly moving
anapole, once the convection term is kept.

This cancellation does not carry over to synthetic motion.  Consider the same
moments $\mathbf d$ and $\boldsymbol\tau$ carried instead by an excitation
pattern that moves through matter at rest.  Such a pattern could be approximated by exciting toroidal or anapole meta-atoms~\cite{KaelbererEtAl2010,ZanganehEtAl2021} in
sequence~\cite{ZhangEtAl2018SpaceTimeCoding}.  The ED and
TD terms are the same as in moving matter, so the TD term again leaves
$2ck_0^2\beta_n\boldsymbol\tau_\perp$.  The convection term is also unchanged,
but the motion-induced moments differ (Sec.~\ref{sec:dipoles}; Appendix~\ref{app:dipoles}).  Unlike the TD carried by matter, which acquires an MD and
an EQ, the TD carried by the pattern acquires neither, and the MD acquired by
its ED is opposite in sign to that in moving matter.  In
Eq.~\eqref{eq:anapole-terms}, the EQ term therefore disappears and the MD term
is halved, so the MD and convection terms now sum to only
$-ck_0^2\beta_n\boldsymbol\tau_\perp$ [Fig.~\ref{fig:anapole}(d)].  Half of
the remainder survives.

To all orders in $\beta$, the multipole decomposition gives
\begin{equation}
 \Pi\widetilde{\mathbf J}=ck(\hat{\mathbf n})\big[k(\hat{\mathbf n})-k_0\big]
 \boldsymbol\tau_\perp=ck_0^2\beta_n\boldsymbol\tau_\perp+O(\beta^2),
 \label{eq:synthetic-anapole}
\end{equation}
in agreement with the exact result.  The factor $k(\hat{\mathbf n})-k_0$
vanishes only in the plane perpendicular to $\mathbf v$, where the Doppler
wavenumber equals $k_0$.  The pattern thus radiates in all other directions
[Fig.~\ref{fig:anapole}(e)].  Its radiation peaks along $+\mathbf v$, and the
forward-to-backward ratio of the transverse-current magnitudes is
$[(1+\beta)/(1-\beta)]^2$.  To leading order in $\beta$, the received power is $\tfrac25\beta^2P_{\rm ED}$, where $P_{\rm ED}$ is the power that
the ED alone radiates at rest [Fig.~\ref{fig:anapole}(b)] (Appendix~\ref{app:dipoles}).  Uniform motion thus leaves an anapole carried by matter nonradiating but
makes a pattern with the same ED and TD radiate.

At rest, the ED and TD terms, $-i\omega_0\Pi\mathbf d$ and
$ck_0^2\Pi\boldsymbol\tau$, have the same angular dependence, so that the
radiation of a source with both depends only on the combination $\mathbf d+ik_0\boldsymbol\tau$.  The anapole, for which this combination vanishes, is therefore nonradiating, and for any
other such source the TD cannot be distinguished from the ED in the radiation.
Whether the TD of a source can be distinguished in its radiation at all has
been debated~\cite{PapasimakisEtAl2016,FernandezCorbatonNanzRockstuhl2017,OspanovaEtAl2023}.  Uniform motion of matter does not change this, whereas acceleration can~\cite{Savinov2018}.  In the laboratory frame, the motion changes the ED and TD terms in the same way.
Through $O(\beta)$, the terms generated by the ED sum to
$-ick_0(\mathbf d_\perp+d_n\boldsymbol\beta_\perp)$ and those generated by the TD
to $ck_0^2(\boldsymbol\tau_\perp+\tau_n\boldsymbol\beta_\perp)$
(Appendix~\ref{app:dipoles}), so the transverse current is
$-ick_0[\mathbf a_\perp+a_n\boldsymbol\beta_\perp]$ with $\mathbf a=\mathbf d+ik_0\boldsymbol\tau$, and the contributions of the
two dipoles keep the same angular dependence.  In synthetic motion, the two terms change differently, because the TD acquires no MD or EQ and the
MD of the ED has the opposite sign.  The multipole decomposition gives
$-ick(\hat{\mathbf n})[\mathbf d_\perp+ik(\hat{\mathbf n})\boldsymbol\tau_\perp]$,
in which the ED and TD terms scale as $k(\hat{\mathbf n})$ and
$k(\hat{\mathbf n})^2$ and so no longer share the same angular dependence.
The ED and TD of the pattern therefore leave separate imprints on its radiation
[Eq.~\eqref{eq:synthetic-anapole}].

\section{Superluminal, nonuniform, and nonrigid motion}
\label{sec:scope}

\begin{figure*}[t]
 \centering
 \includegraphics[width=\textwidth]{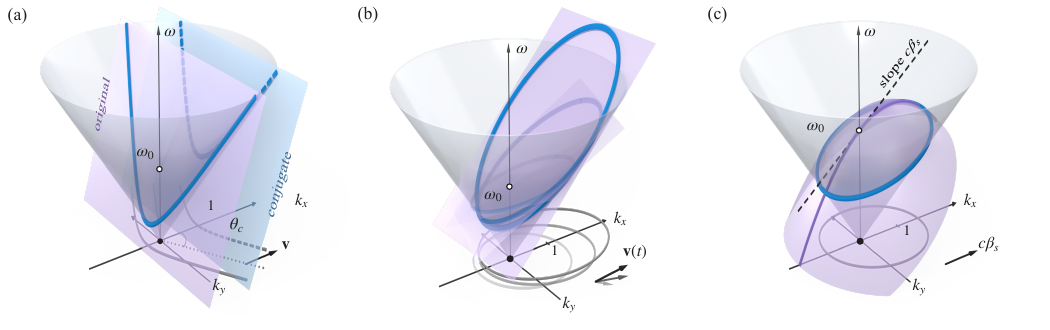}
 \caption{Spectral supports beyond uniform rigid translation, shown as in
Fig.~\ref{fig:support}. (a) Superluminal pattern, $\beta=2$. The supports of
the original and the complex-conjugate components meet the light cone on two
branches, separated by the Cherenkov directions $\theta_c=\arccos(1/\beta)$.
(b) Pattern in nonuniform translation. The instantaneous support
$\omega=\omega_0+\mathbf v(t)\cdot\mathbf k$ turns about the carrier point as
the velocity changes in magnitude and direction (three instants with
$|\boldsymbol\beta|=0.25$, $0.47$, and $0.61$, from light to dark), and each
instantaneous support meets the light cone on its own Doppler ellipse.
(c) Nonrigid pattern with the curved support
$\omega=\omega_0+c\beta_sk_x-\tfrac12c\alpha|\mathbf k|^2$, $\beta_s=0.5$,
$\alpha=0.6/k_0$. The dark curve is its $k_y=0$ profile, and the dashed line is
the tangent of this profile at $\mathbf k=0$, of slope $c\beta_s$.}
 \label{fig:extensions}
\end{figure*}

Sections~\ref{sec:relations}--\ref{sec:anapole} treat a source that translates rigidly at a constant velocity below $c$.  An excitation pattern
need not move this way.  An ionization front driven by a flying focus can move
faster than light~\cite{Palastro2018}, a polarization current can be driven
around a circle~\cite{ArdavanEtAl2004}, and a metasurface can give each
component of its modulation its own frequency~\cite{Wu2025}.
The multipole decomposition still applies.  Given the laboratory-frame
moments, it depends on the motion only through the spectral support
$\omega=\omega_s(\mathbf k)$, because the Doppler wavenumber is where the
support meets the light cone and the convection term follows from the slope
of the support at $\mathbf k=0$.  It therefore suffices to follow how
superluminal, nonuniform, and nonrigid motion change the tilted plane
$\omega=\omega_{s0}+\mathbf v\cdot\mathbf k$ of uniform rigid translation
[Fig.~\ref{fig:support}(b)] and its intersection with the light cone
(Fig.~\ref{fig:extensions}).  We take the three cases in turn.

Only an excitation pattern can move faster than light.  A light spot swept across a distant screen is the classic
example~\cite{BolotovskiiGinzburg1972}.  Such a pattern has no rest frame, and
its moments can be defined only in the laboratory frame.  A multipole decomposition of its radiation must therefore start from the
laboratory-frame current, as in Sec.~\ref{sec:relations}.  The support is still the plane of
Eq.~\eqref{eq:tilted-support}, now steeper than the light cone, and
Eq.~\eqref{eq:doppler-k} gives no positive wavenumber in directions with
$\beta_n>1$.  The radiation in these directions comes from the
complex-conjugate component that every real source contains.  This component
has the carrier frequency $-\omega_0$, and its support meets the light cone at $k_-(\hat{\mathbf n})=k_0/(\beta_n-1)$, an instance of the anomalous Doppler effect~\cite{BolotovskiiGinzburg1972} [Fig.~\ref{fig:extensions}(a)] (Appendix~\ref{app:support}).
Equation~\eqref{eq:moving-multipole} holds for any sign of $\omega_{s0}$, so
it applies to the conjugate component with $\omega_{s0}\to-\omega_0$ and the
moments replaced by their complex conjugates.  The radiation of a superluminal pattern is therefore decomposed into the same
laboratory-frame moments. The Cherenkov directions $\theta_c=\arccos(1/\beta)$,
where the wavenumbers of both components diverge, separate the two branches,
the conjugate component radiating inside the Cherenkov cone and the original
component outside it.  Since the expansion about $\mathbf k=0$ requires
$k(\hat{\mathbf n})a\ll1$ for a source of size $a$, it fails in a band around $\theta_c$ of angular width $2k_0a/\sqrt{\beta^2-1}$
for $k_0a\ll\beta-1$, and for the Gaussian profiles of Appendix~\ref{app:finite}
this band carries most of the radiated power.  In this band, moments of every order contribute comparably, so the
radiation depends on the whole profile of the source.  As $\beta$ falls to 1,
the Cherenkov directions close onto $\mathbf v$.  Below $c$, a pattern meets
the same limit in the forward direction, where
$k(\hat{\mathbf n})=k_0/(1-\beta)$ grows without bound as $\beta\to1$, and the
expansion fails near $1-\beta\approx k_0a$.

A source that moves along a trajectory $\mathbf R(t)$ with a varying velocity
$\mathbf v(t)=\dot{\mathbf R}(t)$ is treated in the same way, with its moments taken about the instantaneous center.  Charge conservation then gives exactly
$M_{ab}+M_{ba}=\dot S_{ab}+v_ap_b+v_bp_a$, where the dot denotes a time
derivative (Appendix~\ref{app:nonuniform}).  This is the relation behind the
EQ and convection terms of Eq.~\eqref{eq:moving-multipole}, with the time
derivative in place of $-i\omega_{s0}$ and the instantaneous velocity in place
of $\mathbf v$, so the convection term follows the velocity at each instant.
The spectrum, however, no longer lies on a single plane.  At each instant the
support is the plane $\omega=\omega_{s0}+\mathbf v(t)\cdot\mathbf k$, and for a
pattern, whose carrier frequency is fixed, this instantaneous support turns about the carrier point $(\mathbf k=0,\ \omega=\omega_0)$ as the velocity changes in magnitude and direction
[Fig.~\ref{fig:extensions}(b)].  The radiation at a given frequency collects
contributions from the whole trajectory through the phase
$e^{i[\omega t-\mathbf k\cdot\mathbf R(t)]}$.  If the source moves many reduced
wavelengths while its velocity changes, this phase is stationary where the
instantaneous support meets the light cone, so the source at each instant radiates at its own Doppler wavenumber.  Faster changes blend these contributions, and the
full phase integral is needed.  In either case the moment relation is exact,
so the multipole terms keep their form, but only in the first case does the source at each instant radiate at its
own Doppler wavenumber.  The rigid pattern carries no net charge, so there is no
acceleration-radiation term associated with a transported net charge.
Because it starts from the laboratory-frame current, the decomposition needs
no rest frame, which also makes it suited to accelerated sources, for which
no single inertial rest frame exists.

Periodic motion is the simplest example.  A source displaced as
$\mathbf R(t)=\xi_0\sin(\Omega t)\hat{\mathbf x}$ radiates at the sidebands
$\omega_\ell=\omega_0+\ell\Omega$ with the Bessel weights
$J_\ell(\omega_\ell\xi_0n_x/c)$, where $\ell$ is an integer and
$n_x=\hat{\mathbf n}\cdot\hat{\mathbf x}$.  For $\omega_0\xi_0/c\gg1$, the
source moves many reduced wavelengths in each period, and these weights trace
the instantaneous Doppler shifts.  When both $\omega_0\xi_0/c$ and the peak speed $\Omega\xi_0/c$ are small, the first sidebands dominate.  Such sidebands are the micro-Doppler signature of
vibrating scatterers~\cite{ChenEtAl2006}, and space--time-coding metasurfaces
produce harmonics at the same frequencies~\cite{ZhangEtAl2018SpaceTimeCoding}.  The decomposition
assigns each sideband to the moments of the source.  For an ED pattern, the
transverse current at $\omega_\ell$ is
\begin{equation}
 \mathbf a_\ell=-i\omega_\ell\,
 J_\ell\!\left(\frac{\omega_\ell\xi_0n_x}{c}\right)\mathbf d_\perp.
 \label{eq:sidebands}
\end{equation}
Here the ED term contributes $-i\omega_0J_\ell\mathbf d_\perp$, and the MD and
convection terms, both proportional to $\mathbf v(t)$, contribute the rest,
$-i\ell\Omega J_\ell\mathbf d_\perp$ (Appendix~\ref{app:nonuniform}).

A pattern whose components carry independently chosen frequencies, as on the
metasurface of Ref.~\cite{Wu2025}, changes its shape as it moves, and its
support is no longer a plane.  As long as each spatial component $\mathbf k$
oscillates at a single frequency $\omega_s(\mathbf k)$, the continuity
relation of Eq.~\eqref{eq:fourier-continuity} holds on the curved support, and
only the matching order by order changes.  We expand
$\omega_s(\mathbf k)=\omega_{s0}+c\boldsymbol\beta_s\cdot\mathbf k
+\tfrac{c}{2}\mathbf k\cdot\mathsf H\cdot\mathbf k+\cdots$ about $\mathbf k=0$.
In the relation for the first current moment, the gradient
$c\boldsymbol\beta_s$ enters exactly as $\mathbf v$ does in
Eq.~\eqref{eq:moment-identity}, and the curvature $\mathsf H$ multiplies only
the total charge, which vanishes for a component oscillating at $\omega_{s0}\neq0$.  Equation~\eqref{eq:moving-multipole} thus holds at the order of the MD, EQ,
and convection terms, with $\mathbf v$ replaced by $c\boldsymbol\beta_s$ and
evaluated where the curved support meets the light cone [Fig.~\ref{fig:extensions}(c)].  The curvature first enters at
the order of the TD, where it adds a term proportional to $\mathsf H$ and
$\mathbf p$ (Appendix~\ref{app:continuity}).  The gradient
$c\boldsymbol\beta_s$ is the group velocity of the pattern, as the slope of
the support is for space--time wave packets in free
space~\cite{SaariReivelt2004,YessenovEtAl2019,KondakciAbouraddy2019}.  The multipole decomposition is therefore determined by the spectral support
rather than by the motion itself.  Motion is one way to tilt the support, and any other way of
making the frequency depend on the wavevector enters the decomposition through
the same support, its slope setting the convection term and its intersection
with the light cone setting the Doppler wavenumber.

\section{Conclusion}
\label{sec:conclusion}

We have formulated the multipole decomposition of moving sources in terms of
laboratory-frame moments.  Compared with a source at rest, two things change.
Each multipole radiates at the Doppler wavenumber of its direction, and a
convection term, absent for stationary sources, is added.  The decomposition
shows that the same dipole radiates differently in moving matter and in
synthetic motion, because the two kinds of motion induce different laboratory-frame moments.  For an anapole, the difference between these induced moments decides whether it radiates.  In moving matter, the motion changes the ED and TD terms in the same way, and the anapole remains nonradiating.  A pattern with the same ED and TD, by contrast,
radiates, and its ED and TD leave separate imprints on the radiation.  For a
moving pattern, the condition for being nonradiating must therefore be
reconsidered.

Because the motion enters only through the spectral support, the
decomposition holds equally for superluminal, nonuniform, and nonrigid motion,
and for any source whose spectrum lies on a smooth
support $\omega=\omega_s(\mathbf k)$. It therefore serves accelerated sources,
for which no single rest frame exists, and localized patterns in synthetic
motion, which space--time-modulated metamaterials now realize in translation
and in rotation~\cite{Huidobro2019,Galiffi2022,HarwoodEtAl2025,Liberal2026,Nasari2026}.
In both kinds of motion, the radiation is assigned to the same set of multipole moments as for a stationary source, so that the
multipolar toolkit, from identifying multipolar resonances to designing
Kerker-type and anapole states, applies to moving sources as well.  The moments and the spectral support
together set the radiation and become parameters for its design.  For moving
matter these include the velocity and the trajectory, and for a pattern the
support itself can be shaped.  The multipole decomposition, so far a tool for
stationary sources, thus extends to moving matter and to patterns in synthetic
motion and opens the way from understanding their radiation to designing it.

\begin{acknowledgments}
This work was supported by the National Natural Science Foundation of China
(No. 62201541).
\end{acknowledgments}

\noindent\textit{Data availability.}
The data and code supporting the findings of this study are available from the
corresponding author upon reasonable request.

\appendix

\makeatletter
\renewcommand\subsection{\@startsection{subsection}{2}{\z@}{10pt plus 2pt minus 2pt}{5pt}{\normalfont\small\bfseries\centering}}
\makeatother

\section{Sources and their radiation}
\label{app:sources}

The appendices derive the relations used in the main text and give the
source models behind its exact results.  Appendix~\ref{app:sources} fixes the
conventions and derives the spectral support, the Doppler wavenumber, and the
received power of a translating source.  Appendix~\ref{app:continuity}
derives the continuity relations and the multipole decomposition, Eqs.~\eqref{eq:moment-identity}
and \eqref{eq:moving-multipole}, at all orders and on curved supports.  Appendix~\ref{app:dipoles}
gives the exact laboratory currents of the elementary dipoles and of the
anapole, Appendix~\ref{app:nonuniform} treats nonuniform translation, and
Appendix~\ref{app:finite} the finite size of the source.  A remainder
$O(\beta^2)$ scales at least quadratically in $\beta$ with all other
variables fixed, and results stated without a remainder are exact within the
stated source model.

\subsection{Conventions and moments}

We use Gaussian units and the time factor $e^{-i\omega t}$.  Cartesian
indices $a,b,j\in\{1,2,3\}$ are summed when repeated, ${}^{\mathsf T}$
denotes the transpose, $I$ the $3\times3$ identity, and products such as
$\mathbf a\cdot\mathbf b$ are bilinear also for complex vectors.  The
spatial and space--time transforms are
\begin{equation}
 \begin{gathered}
 \widetilde f(\mathbf k)=\int f(\mathbf r)e^{-i\mathbf k\cdot\mathbf r}d^3r,
 \\
 \widetilde{\mathbf J}(\omega,\mathbf k)
 =\int\! dt\!\int\! d^3r\,e^{i\omega t-i\mathbf k\cdot\mathbf r}\mathbf J.
 \end{gathered}
 \label{eq:A-fourier}
\end{equation}
For a unit vector $\hat{\mathbf n}$,
$\Pi(\hat{\mathbf n})=I-\hat{\mathbf n}\hat{\mathbf n}^{\mathsf T}$, and
$a_n=\mathbf a\cdot\hat{\mathbf n}$, $\mathbf a_\perp=\Pi(\hat{\mathbf n})\mathbf a$
for any vector $\mathbf a$.  A source moves at $\mathbf v=c\boldsymbol\beta$,
$\hat{\mathbf v}=\mathbf v/|\mathbf v|$, and $\gamma=(1-\beta^2)^{-1/2}$ for
$\beta<1$.  The polar angle $\theta$ is measured from $\mathbf v$, and
forward and backward refer to $\hat{\mathbf n}=\pm\hat{\mathbf v}$.  Rest-frame
quantities of moving matter carry a prime.  Point sources are
understood as distributions.

The moments are integrals over the source at a fixed laboratory time, with
$\mathbf r$ measured from the instantaneous center of the source,
\begin{equation}
 \begin{gathered}
 q=\!\int\!\rho\,d^3r,\quad p_a=\!\int\! r_a\rho\,d^3r,\quad
 S_{ab}=\!\int\! r_ar_b\rho\,d^3r,\\
 \mathsf Q=3\mathsf S-(\mathrm{tr}\,\mathsf S)I,\quad
 M_{ab}=\!\int\! r_bJ_a\,d^3r,\\
 \mathbf m=\frac{1}{2c}\!\int\!\mathbf r\times\mathbf J\,d^3r,\\
 \mathbf T=\frac{1}{10c}\int\big[\mathbf r(\mathbf r\cdot\mathbf J)
 -2r^2\mathbf J\big]d^3r,
 \end{gathered}
 \label{eq:A-moments}
\end{equation}
namely the total charge, the ED, the second charge moment, the traceless EQ,
the first current moment, the MD, and the TD.  In Fourier space,
\begin{equation}
 \begin{gathered}
 q=\widetilde\rho(\mathbf 0),\quad p_a=i\partial_a\widetilde\rho(\mathbf 0),
 \quad S_{ab}=-\partial_a\partial_b\widetilde\rho(\mathbf 0),\\
 M_{ab}=i\partial_b\widetilde J_a(\mathbf 0),\quad
 m_a=\frac{i}{2c}\epsilon_{abj}\partial_b\widetilde J_j(\mathbf 0),\\
 T_a=-\frac{1}{10c}\big(\partial_a\partial_b\widetilde J_b
 -2\partial_b\partial_b\widetilde J_a\big)(\mathbf 0),
 \end{gathered}
 \label{eq:A-fourier-moments}
\end{equation}
with $\partial_a=\partial/\partial k_a$.  Moving the origin to $\mathbf r_0$
multiplies the amplitudes by $e^{i\mathbf k\cdot\mathbf r_0}$, which leaves
$|\Pi\widetilde{\mathbf J}|$ and the radiated power unchanged but changes the moments to
\begin{equation}
 \mathbf p_{\mathbf r_0}=\mathbf p-q\mathbf r_0,\qquad
 \mathbf m_{\mathbf r_0}=\mathbf m+\frac{i\omega_{s0}}{2c}\mathbf r_0\times\mathbf p
 \quad(q=0),
 \label{eq:A-origin}
\end{equation}
where $\omega_{s0}$ is the carrier frequency (Sec.~\ref{app:support}).
About the instantaneous center, the moments of a translating source at
different times differ only by the factor $e^{-i\omega_{s0}t}$, whereas about
a fixed origin they change with the position of the source.  The individual multipole terms therefore depend on the choice of origin, here the instantaneous center, whereas the modulus of their sum does not.

\subsection{Elementary sources}

At rest, an ED $\mathbf d$, an MD $\boldsymbol\mu$, and a TD $\boldsymbol\tau$
oscillating at $\omega_0=ck_0$ are represented by the conserved point sources~\cite{NemkovBasharinFedotov2017}
\begin{equation}
 \begin{aligned}
 \rho_d&=-(\mathbf d\cdot\nabla)\delta^3(\mathbf r),&
 \mathbf J_d&=-i\omega_0\mathbf d\,\delta^3(\mathbf r),\\
 \mathbf J_\mu&=c\nabla\times[\boldsymbol\mu\,\delta^3(\mathbf r)],&
 \mathbf J_\tau&=c\nabla\times\nabla\times[\boldsymbol\tau\,\delta^3(\mathbf r)],
 \end{aligned}
 \label{eq:A-rest-sources}
\end{equation}
with $\rho_\mu=\rho_\tau=0$, each multiplied by $e^{-i\omega_0t}$.  Their
amplitudes are $\widetilde{\mathbf J}_d=-i\omega_0\mathbf d$,
$\widetilde{\mathbf J}_\mu=ic\mathbf k\times\boldsymbol\mu$, and
$\widetilde{\mathbf J}_\tau=c\,\mathbf w(\mathbf k)$ with
$\mathbf w(\mathbf k)=|\mathbf k|^2\boldsymbol\tau-\mathbf k(\mathbf k\cdot\boldsymbol\tau)$,
and Eq.~\eqref{eq:A-fourier-moments} gives $\mathbf p=\mathbf d$,
$\mathbf m=\boldsymbol\mu$, $\mathbf T=\boldsymbol\tau$, and no EQ.  The
anapole combines an ED and a TD with
\begin{equation}
 \mathbf d=-ik_0\boldsymbol\tau,\qquad\text{so that}\qquad
 \mathbf p+ik_0\mathbf T=0.
 \label{eq:A-anapole}
\end{equation}
The sign belongs to the time factor $e^{-i\omega_0t}$ and to the definition
of $\mathbf T$ in Eq.~\eqref{eq:A-moments}.

Every source can be written as
\begin{equation}
 \rho=-\nabla\cdot\boldsymbol{\mathcal P},\qquad
 \mathbf J=\partial_t\boldsymbol{\mathcal P}+c\nabla\times\boldsymbol{\mathcal M},
 \label{eq:A-PM}
\end{equation}
which conserves charge identically.  In moving matter, the sources of
Eq.~\eqref{eq:A-rest-sources} describe the matter in its rest frame, and the
laboratory source is their Lorentz transform (Appendix~\ref{app:dipoles}).
In synthetic motion, the matter stays at rest, and the pattern of
$\boldsymbol{\mathcal P}$ and $\boldsymbol{\mathcal M}$, which includes the
full response of the matter, is prescribed in the laboratory and radiates as a
source in vacuum.  With $\delta_v=\delta^3(\mathbf r-\mathbf vt)$, the
synthetic ED, MD, and TD have, respectively,
\begin{equation}
 \boldsymbol{\mathcal P}=\mathbf d\,\delta_v,\qquad
 \boldsymbol{\mathcal M}=\boldsymbol\mu\,\delta_v,\qquad
 \boldsymbol{\mathcal M}=\nabla\times(\boldsymbol\tau\delta_v),
 \label{eq:A-synthetic}
\end{equation}
each multiplied by $e^{-i\omega_0t}$ and with the other density zero, where $\omega_0$ is the excitation frequency and $\partial_t$ in
Eq.~\eqref{eq:A-PM} acts on $\delta_v$ as well.  Only the pattern moves, so
$|\mathbf v|$ may exceed $c$.

\subsection{Spectral support and Doppler wavenumber}
\label{app:support}

A source translating at $\mathbf v$ while oscillating at $\omega_{s0}$ obeys
$\mathbf J(\mathbf r,t)=\mathbf J(\mathbf r-\mathbf vt,0)e^{-i\omega_{s0}t}$,
and likewise for $\rho$, so that
\begin{equation}
 \widetilde{\mathbf J}(\omega,\mathbf k)=2\pi\delta\big(\omega-\omega_s(\mathbf k)\big)
 \widetilde{\mathbf J}(\mathbf k),\qquad
 \omega_s(\mathbf k)=\omega_{s0}+\mathbf v\cdot\mathbf k,
 \label{eq:A-support}
\end{equation}
which is the content of Eqs.~\eqref{eq:translating-source} and \eqref{eq:tilted-support}.  In synthetic
motion $\omega_{s0}=\omega_0$.  In moving matter, the Lorentz transformation
$\omega'=\gamma(\omega-\mathbf v\cdot\mathbf k)$ maps the rest-frame support
$\omega'=\omega_0$ onto $\omega=\omega_0/\gamma+\mathbf v\cdot\mathbf k$, so
that $\omega_{s0}=\omega_0/\gamma$, and
$\delta[\gamma(\omega-\mathbf v\cdot\mathbf k)-\omega_0]
=\gamma^{-1}\delta(\omega-\omega_s)$ gives every laboratory amplitude of
moving matter the factor $1/\gamma$.  On this support the rest-frame
wavevector is
\begin{equation}
 \mathbf k'=\mathsf B\mathbf k-k_0\boldsymbol\beta,\qquad
 \mathsf B=I-\frac{\gamma}{\gamma+1}\boldsymbol\beta\boldsymbol\beta^{\mathsf T},
 \label{eq:A-kprime}
\end{equation}
where $\mathsf B$ multiplies the component along $\mathbf v$ by $1/\gamma$.

A plane-wave component radiates only on the light cone $\omega=c|\mathbf k|$.
For $\mathbf k=k\hat{\mathbf n}$, Eq.~\eqref{eq:A-support} gives, for
$\beta_n<1$, the Doppler wavenumber
\begin{equation}
 k(\hat{\mathbf n})=\frac{\omega_{s0}}{c(1-\beta_n)}=k_0(1+\beta_n)+O(\beta^2),
 \label{eq:A-doppler}
\end{equation}
which is $k_0/[\gamma(1-\beta_n)]$ in moving matter and $k_0/(1-\beta_n)$ in
synthetic motion.  The radiating wavevectors form a prolate spheroid
about $\mathbf v$ with a focus at $\mathbf k=0$ [Fig.~\ref{fig:support}(b)].  Its meridian is an ellipse centered at
$k_\parallel=\beta\gamma k_0$ with semiaxes $\gamma k_0$ along $\mathbf v$
and $k_0$ across it in moving matter, and centered at $\beta\gamma^2k_0$ with
semiaxes $\gamma^2k_0$ and $\gamma k_0$ in synthetic motion.

A real source also contains the complex-conjugate component, with carrier
$-\omega_{s0}$ and amplitudes $\widetilde\rho_-(\mathbf k)=[\widetilde\rho(-\mathbf k)]^*$
and $\widetilde{\mathbf J}_-(\mathbf k)=[\widetilde{\mathbf J}(-\mathbf k)]^*$,
whose moments are the complex conjugates of those of the original component.
Its support is $\omega=-\omega_{s0}+\mathbf v\cdot\mathbf k$.  For
$\beta_n>1$, which only synthetic motion reaches, the original component
does not meet the positive-frequency light cone, and the conjugate component
radiates at
\begin{equation}
 k_-(\hat{\mathbf n})=\frac{\omega_{s0}}{c(\beta_n-1)},
 \label{eq:A-conjugate}
\end{equation}
the anomalous Doppler branch.  At $\beta_n=1$, the Cherenkov directions
$\theta_c=\arccos(1/\beta)$, neither component meets the light cone at a
finite wavenumber.  For $\beta_n<1$, the conjugate component lies on the
negative-frequency light cone and supplies the complex conjugate of the
original field.

\subsection{Far field and received power}
\label{app:power}

In the far field at distance $L$, each component of angular frequency
$\omega$ and wavenumber $k=\omega/c$ gives~\cite{Jackson1999}
\begin{equation}
 \mathbf E=\frac{ik\,e^{ikL}}{cL}\,\Pi(\hat{\mathbf n})
 \widetilde{\mathbf J}(\omega,k\hat{\mathbf n}),
 \label{eq:A-farfield}
\end{equation}
with the magnetic field $\hat{\mathbf n}\times\mathbf E$,
where charge conservation, $\mathbf k\cdot\widetilde{\mathbf J}=\omega\widetilde\rho$,
makes the scalar potential cancel the longitudinal part of the vector
potential.  The charge density thus enters only through continuity, and the
radiation is fixed by the transverse current alone~\cite{Jackson1999,DevaneyWolf1973}.  For the
translating source of Eq.~\eqref{eq:A-support}, the inverse time transform
carries the Jacobian $|1-\beta_n|^{-1}$ of the root $\omega_n=ck(\hat{\mathbf n})$,
so that the power per solid angle measured per reception time is
\begin{equation}
 \frac{dP_{\rm rec}}{d\Omega}
 =\frac{ck(\hat{\mathbf n})^4}{8\pi\omega_{s0}^2}
 \big|\Pi(\hat{\mathbf n})\widetilde{\mathbf J}\big(k(\hat{\mathbf n})\hat{\mathbf n}\big)\big|^2 .
 \label{eq:A-power}
\end{equation}
For $\beta_n>1$ the conjugate component gives the same expression with $k_-$
and $\widetilde{\mathbf J}_-$.  The prefactor depends on the direction and on
the carrier but not on how the current is expanded, so every approximation to
the current of one source is multiplied by the same prefactor, which carries
the forward beaming.  Its forward-to-backward ratio is $\eta^4$, with
\begin{equation}
 \eta=\frac{1+\beta}{|1-\beta|}.
 \label{eq:A-eta}
\end{equation}
For moving matter, Eq.~\eqref{eq:A-power} equals $D^4\,dP'/d\Omega'$ with the
Doppler factor $D=[\gamma(1-\beta_n)]^{-1}$ and the rest-frame power in the
aberrated direction~\cite{Jackson1999}, and for $\mathbf v=0$ and an ED it
integrates to the Larmor power $\omega_0^4|\mathbf d|^2/(3c^3)$.

The continuity relations of Appendix~\ref{app:continuity} do not involve the
light cone.  In a homogeneous, nondispersive, nonmagnetic background of index $n_b$
at rest, the light cone becomes $\omega=c|\mathbf k|/n_b$, the Doppler
wavenumber $n_b\omega_{s0}/[c(1-n_b\beta_n)]$, and the Cherenkov condition
$n_b\beta_n=1$, while the multipole expansion of the transverse current is
unchanged.

\section{Continuity relations and the multipole decomposition}
\label{app:continuity}

\subsection{Continuity on a smooth spectral support}

We derive the relations on a general smooth support
$\omega=\omega_s(\mathbf k)$ and specialize to translation afterward.  With
the common distribution on the support removed, as in
Eq.~\eqref{eq:A-support}, charge conservation reads
\begin{equation}
 \mathbf k\cdot\widetilde{\mathbf J}(\mathbf k)=\omega_s(\mathbf k)\,\widetilde\rho(\mathbf k).
 \label{eq:B-continuity}
\end{equation}
We define the intercept, the dimensionless gradient, and the curvature of the
support,
\begin{equation}
 \omega_{s0}=\omega_s(\mathbf 0),\quad
 \beta_{s,a}=\frac1c\partial_a\omega_s(\mathbf 0),\quad
 H_{ab}=\frac1c\partial_a\partial_b\omega_s(\mathbf 0),
 \label{eq:B-support-coeff}
\end{equation}
so that $c\boldsymbol\beta_s=\mathbf v$ and $\mathsf H=0$ for translation.
Matching the terms of degree zero, one, and two in $\mathbf k$ in
Eq.~\eqref{eq:B-continuity} gives
\begin{equation}
 \begin{gathered}
 \omega_{s0}q=0,\qquad
 \widetilde J_a(\mathbf 0)=c\beta_{s,a}q-i\omega_{s0}p_a,\\
 M_{ab}+M_{ba}=-i\omega_{s0}S_{ab}+c(\beta_{s,a}p_b+\beta_{s,b}p_a)+icH_{ab}q.
 \end{gathered}
 \label{eq:B-first}
\end{equation}
A component with $\omega_{s0}\neq0$ is therefore neutral, its net current is
that of its oscillating ED, and the symmetric part of its first current
moment is fixed, while the antisymmetric part, which carries the MD, is free.
The relations involve only the signed $\omega_{s0}$ and hold for the
conjugate component with $\omega_{s0}\to-\omega_{s0}$.  They relate Taylor
coefficients and hold for extended sources as well.  For instance,
$\widetilde\rho=-i(\mathbf p\cdot\mathbf k)F$ and
$\widetilde{\mathbf J}=-i\omega_s(\mathbf k)\mathbf pF$ with the form factor
$F=e^{-|\mathbf k|^2a^2/2}$ satisfy Eq.~\eqref{eq:B-continuity} for any smooth
$\omega_s$ and reproduce Eq.~\eqref{eq:B-first}.

\subsection{The first current moment}

The current expands as
$\widetilde J_a(\mathbf k)=\widetilde J_a(\mathbf 0)-iM_{ab}k_b
-\tfrac12N_{abj}k_bk_j+\cdots$, with $N_{abj}=\int r_br_jJ_a\,d^3r$.  The
antisymmetric part of $M_{ab}$ gives the MD term,
$-\tfrac i2(M_{ab}-M_{ba})k_b=ic(\mathbf k\times\mathbf m)_a$, which is
already transverse.  With $\Pi=\Pi(\hat{\mathbf k})$, the symmetric part
follows from Eq.~\eqref{eq:B-first},
\begin{equation}
 \begin{aligned}
 -\frac i2\Pi\big[(\mathsf M+\mathsf M^{\mathsf T})\mathbf k\big]
 ={}&-\frac{\omega_{s0}}{6}\Pi(\mathsf Q\mathbf k)+\frac{cq}{2}\Pi(\mathsf H\mathbf k)\\
 &-\frac{ic}{2}\Pi\big[\boldsymbol\beta_s(\mathbf p\cdot\mathbf k)
 +\mathbf p(\boldsymbol\beta_s\cdot\mathbf k)\big],
 \end{aligned}
 \label{eq:B-symmetric}
\end{equation}
where $\Pi\mathbf k=0$ has removed the trace of $\mathsf S$.  The first term is
the EQ term and the last the convection term.  The second, the curvature
term at this order, is proportional to $q$ and vanishes for
$\omega_{s0}\neq0$.  A charged component with $\omega_{s0}=0$ retains it, as
the example $\omega_s=\tfrac c2\mathbf k^{\mathsf T}\mathsf H\mathbf k$,
$\widetilde\rho=q$, $\widetilde{\mathbf J}=\tfrac{cq}{2}\mathsf H\mathbf k$
shows.

\subsection{Translating source}

For translation, Eq.~\eqref{eq:B-first} becomes, for any $\beta$,
\begin{equation}
 M_{ab}+M_{ba}=-i\omega_{s0}S_{ab}+v_ap_b+v_bp_a,
 \label{eq:B-translating}
\end{equation}
which is Eq.~\eqref{eq:moment-identity}.  The terms of degree zero and one, and the
TD term of degree two (Sec.~\ref{app:second}), give Eq.~\eqref{eq:moving-multipole},
\begin{equation}
 \begin{aligned}
 \Pi\widetilde{\mathbf J}={}&-i\omega_{s0}\Pi\mathbf p+c|\mathbf k|^2\Pi\mathbf T
 +ic\mathbf k\times\mathbf m\\
 &-\frac{\omega_{s0}}{6}\Pi(\mathsf Q\mathbf k)
 +\boldsymbol{\mathcal A}_v(\mathbf k)+\cdots,
 \end{aligned}
 \label{eq:B-eq8}
\end{equation}
with the convection term
\begin{equation}
 \boldsymbol{\mathcal A}_v(\mathbf k)=-\frac i2\Pi\big[\mathbf v(\mathbf p\cdot\mathbf k)
 +\mathbf p(\mathbf v\cdot\mathbf k)\big],
 \label{eq:B-convection}
\end{equation}
labeled ``Conv.'' in Figs.~\ref{fig:anapole}(c) and \ref{fig:anapole}(d).  The radiation
follows at $\mathbf k=k(\hat{\mathbf n})\hat{\mathbf n}$.  For $\mathbf v=0$
and $\mathbf k=k_0\hat{\mathbf n}$, Eq.~\eqref{eq:B-eq8} reduces to
\begin{equation}
 \Pi\widetilde{\mathbf J}=-i\omega_0\Pi(\mathbf p+ik_0\mathbf T)
 +ick_0\hat{\mathbf n}\times\mathbf m-\frac{\omega_0k_0}{6}\Pi(\mathsf Q\hat{\mathbf n}),
 \label{eq:B-static}
\end{equation}
the relations of a stationary source~\cite{RadescuVaman2002,AlaeeRockstuhlFernandezCorbaton2019},
in which the ED and the TD enter only through $\mathbf p+ik_0\mathbf T$.  The
conventional relations are the first four terms of Eq.~\eqref{eq:B-eq8} with the laboratory-frame moments, evaluated at $\mathbf k=(\omega_{s0}/c)\hat{\mathbf n}$,
where $\omega_{s0}/c=k_0/\gamma=k_0+O(\beta^2)$ for moving matter.
The same four terms evaluated at $\mathbf k=k(\hat{\mathbf n})\hat{\mathbf n}$
are the Doppler-shifted relations.

\subsection{The second current moment}
\label{app:second}

The second current moment contributes $-\tfrac12N_{abj}k_bk_j$ to
$\widetilde J_a$.  Splitting $N_{abj}$ into its part antisymmetric in $a$ and
$b$, which involves $\mathbf r\times\mathbf J$, and its totally symmetric part
$\bar N_{abj}=N_{(abj)}$, one finds the transverse current of degree two,
\begin{equation}
 \begin{aligned}
 -\frac12\Pi\big(N_{\cdot bj}k_bk_j\big)
 ={}&c|\mathbf k|^2\Pi\mathbf T\\
 &+\frac c2\mathbf k\times(\mathsf M_2\mathbf k)
 -\frac12\Pi\big(\bar N^{\rm STF}_{\cdot bj}k_bk_j\big),
 \end{aligned}
 \label{eq:B-second}
\end{equation}
where
$(\mathsf M_2)_{ab}=\frac{1}{3c}\int[r_a(\mathbf r\times\mathbf J)_b
+r_b(\mathbf r\times\mathbf J)_a]d^3r$ is the traceless magnetic quadrupole
(MQ) and $\bar N^{\rm STF}$ the symmetric traceless part of $\bar N$.  The vector
$\mathbf u=\int[r^2\mathbf J-\mathbf r(\mathbf r\cdot\mathbf J)]d^3r$ of the
antisymmetric part and the trace
$\mathbf t=\tfrac13\int[r^2\mathbf J+2\mathbf r(\mathbf r\cdot\mathbf J)]d^3r$
of the symmetric part contribute $-\tfrac16|\mathbf k|^2\Pi\mathbf u$ and
$-\tfrac1{10}|\mathbf k|^2\Pi\mathbf t$, which together give
$c|\mathbf k|^2\Pi\mathbf T$.  The TD is thus defined by the current alone and
does not involve continuity.  The MQ and the symmetric traceless part are
among the terms not written out in Eq.~\eqref{eq:B-eq8}.

\subsection{All orders}

Continuity fixes the totally symmetric part of every current moment.  With
the charge and current moments
$C_{a_1\cdots a_m}=\int r_{a_1}\cdots r_{a_m}\rho\,d^3r$ and
$I_{a_0|a_1\cdots a_n}=\int r_{a_1}\cdots r_{a_n}J_{a_0}\,d^3r$, matching the
terms of degree $n+1$ in Eq.~\eqref{eq:B-continuity} on the plane
$\omega_s=\omega_{s0}+\mathbf v\cdot\mathbf k$ gives, exactly in $\mathbf v$,
\begin{equation}
 (n+1)I_{(a_0|a_1\cdots a_n)}=-i\omega_{s0}C_{a_0\cdots a_n}
 +\sum_{\sigma=0}^{n}v_{a_\sigma}C_{a_0\cdots\widehat{a_\sigma}\cdots a_n},
 \label{eq:B-all-orders}
\end{equation}
where the hat omits an index.  For $n=0$ and $1$ this is
Eqs.~\eqref{eq:B-first} and \eqref{eq:B-translating}, and for $n=2$ it reads
$3\bar N_{abj}=-i\omega_{s0}C_{abj}+v_aS_{bj}+v_bS_{ja}+v_jS_{ab}$.  Inserted
into Eq.~\eqref{eq:B-second}, it turns the symmetric traceless part into an
electric-octupole term and a term in which the second charge moment is
transported at $\mathbf v$.  Equation~\eqref{eq:B-eq8} completed through the
second current moment is then
\begin{equation}
 \begin{aligned}
 \Pi\widetilde{\mathbf J}={}&[\text{Eq.~\eqref{eq:B-eq8}}]
 +\frac c2\mathbf k\times(\mathsf M_2\mathbf k)\\
 &+\frac{i\omega_{s0}}{6}\Pi\big(C^{\rm STF}_{\cdot bj}k_bk_j\big)\\
 &-\frac{1}{18}\Pi\Big[\mathbf v(\mathbf k\cdot\mathsf Q\mathbf k)
 \\
 &\qquad+2(\mathbf v\cdot\mathbf k)\mathsf Q\mathbf k
 -\frac{2|\mathbf k|^2}{5}\mathsf Q\mathbf v\Big]
 +O(|\mathbf k|^3),
 \end{aligned}
 \label{eq:B-completed}
\end{equation}
where $O(|\mathbf k|^3)$ refers to the spatial expansion, not to $\beta$.  The
added terms are the MQ, the electric octupole, and the convection term
carried by the EQ, in the same way as the first current moment yields the MD,
the EQ, and the convection term carried by the ED.  A point TD carried by
moving matter, and an anapole formed from it and an ED, have laboratory
currents that are polynomials of degree two in $\mathbf k$, with vanishing
third charge moment and higher current moments, so Eq.~\eqref{eq:B-completed}
reproduces their currents exactly for any $\beta<1$
(Appendix~\ref{app:dipoles}).

\subsection{Curved supports}

On a curved support, Eq.~\eqref{eq:B-symmetric} shows that, through the first
current moment, the gradient $c\boldsymbol\beta_s$ enters exactly as
$\mathbf v$ does and the curvature enters only through $q$.
Equation~\eqref{eq:B-eq8} therefore holds at this order with
$\mathbf v\to c\boldsymbol\beta_s$, evaluated at the wavenumber fixed by
$ck(\hat{\mathbf n})=\omega_s\big(k(\hat{\mathbf n})\hat{\mathbf n}\big)$.  At
the next order the curvature couples to the ED.  Matching the terms of degree
three in Eq.~\eqref{eq:B-continuity} gives
\begin{equation}
 \begin{aligned}
 3\bar N_{abj}={}&-i\omega_{s0}C_{abj}
 +c(\beta_{s,a}S_{bj}+\beta_{s,b}S_{ja}+\beta_{s,j}S_{ab})\\
 &+ic(H_{ab}p_j+H_{bj}p_a+H_{ja}p_b)-cqH_{abj},
 \end{aligned}
 \label{eq:B-curved}
\end{equation}
with $H_{abj}=c^{-1}\partial_a\partial_b\partial_j\omega_s(\mathbf 0)$.  The terms in $H_{ab}p_j$ are the curvature contribution at the order of the TD
term, and Eq.~\eqref{eq:B-curved} reduces to the case $n=2$ of
Eq.~\eqref{eq:B-all-orders} on a plane. Through the symmetric traceless part
in Eq.~\eqref{eq:B-second}, they add to the transverse current
\begin{equation}
 \begin{aligned}
 \Delta\Pi\widetilde{\mathbf J}_{\mathsf H}={}&-\frac{ic}{6}\Pi\Big[
 2(\mathbf p\cdot\mathbf k)\mathsf H\mathbf k
 +(\mathbf k\cdot\mathsf H\mathbf k)\mathbf p\\
 &-\frac{|\mathbf k|^2}{5}\big(2\mathsf H\mathbf p
 +(\mathrm{tr}\,\mathsf H)\mathbf p\big)\Big],
 \end{aligned}
 \label{eq:B-curvature-term}
\end{equation}
whereas their trace belongs to the TD, which is defined by the current.
Where the support meets the light cone at several wavenumbers in one
direction, each root contributes, with the Jacobian
$|1-\hat{\mathbf n}\cdot\nabla_{\mathbf k}\omega_s/c|^{-1}$ in place of
$|1-\beta_n|^{-1}$ in Eq.~\eqref{eq:A-power}, and the conjugate component,
whose support is $\omega=-\omega_s(-\mathbf k)$, is treated as in
Appendix~\ref{app:support}.  The gradient $c\boldsymbol\beta_s$
describes the spectrum, and for space--time wave packets it is the group
velocity~\cite{YessenovEtAl2019}.

\section{Elementary dipoles and the anapole}
\label{app:dipoles}

\subsection{Laboratory currents of moving matter}

The laboratory four-current $J^\mu=(c\rho,\mathbf J)$ of moving matter is the
Lorentz transform of the rest-frame one, and the space--time transforms obey
$\widetilde J^\mu(\omega,\mathbf k)=\Lambda^\mu{}_\nu\widetilde J'^\nu(\omega',\mathbf k')$.
For a rest-frame source oscillating at $\omega_0$, the support factor of
Sec.~\ref{app:support} gives the laboratory amplitudes, for any $\beta<1$,
\begin{equation}
 \begin{aligned}
 c\widetilde\rho(\mathbf k)&=c\widetilde\rho'(\mathbf k')
 +\boldsymbol\beta\cdot\widetilde{\mathbf J}'(\mathbf k'),\\
 \widetilde{\mathbf J}(\mathbf k)&=\gamma^{-1}\mathsf B^{-1}\widetilde{\mathbf J}'(\mathbf k')
 +c\widetilde\rho'(\mathbf k')\boldsymbol\beta,
 \end{aligned}
 \label{eq:C-boost}
\end{equation}
with $\mathbf k'$ of Eq.~\eqref{eq:A-kprime} and
$\mathsf B^{-1}=I+(\gamma-1)\hat{\mathbf v}\hat{\mathbf v}^{\mathsf T}$.  These
are exact laboratory currents.  At $\mathbf k=k(\hat{\mathbf n})\hat{\mathbf n}$
they give the exact patterns of Fig.~\ref{fig:dipoles} through Eq.~\eqref{eq:A-power} without
a multipole expansion.

\subsection{The ED and the MD in moving matter}

For the ED and the MD of Eq.~\eqref{eq:A-rest-sources},
Eq.~\eqref{eq:C-boost} gives
\begin{equation}
 \begin{aligned}
 \text{ED:}\quad&\widetilde{\mathbf J}=-i\omega_s(\mathbf k)\mathsf B\mathbf d
 -ic\mathbf k\times(\boldsymbol\beta\times\mathbf d),\\
 &\widetilde\rho=-i\mathbf k\cdot\mathsf B\mathbf d,\\
 \text{MD:}\quad&\widetilde{\mathbf J}=-i\omega_s(\mathbf k)(\boldsymbol\beta\times\boldsymbol\mu)
 +ic\mathbf k\times\mathsf B\boldsymbol\mu,\\
 &\widetilde\rho=-i\mathbf k\cdot(\boldsymbol\beta\times\boldsymbol\mu).
 \end{aligned}
 \label{eq:C-ED-MD}
\end{equation}
Both have the form $\widetilde{\mathbf J}=-i\omega_s\widetilde{\boldsymbol{\mathcal P}}
+ic\mathbf k\times\widetilde{\boldsymbol{\mathcal M}}$ of Eq.~\eqref{eq:A-PM},
with $\boldsymbol{\mathcal P}=\mathsf B\mathbf d\,\delta_ve^{-i\omega_{s0}t}$ and
$\boldsymbol{\mathcal M}=(\mathbf d\times\boldsymbol\beta)\delta_ve^{-i\omega_{s0}t}$
for the ED, which also follow from the transformation of polarization and
magnetization with the Lorentz contraction and time dilation.
Equation~\eqref{eq:A-fourier-moments} gives the laboratory-frame moments, for any
$\beta<1$,
\begin{equation}
 \begin{aligned}
 \text{ED:}\quad&\mathbf p=\mathsf B\mathbf d,\quad
 \mathbf m=-\tfrac12\boldsymbol\beta\times\mathbf d,\\
 \text{MD:}\quad&\mathbf p=\boldsymbol\beta\times\boldsymbol\mu,\quad
 \mathbf m=\mathsf B\boldsymbol\mu+\tfrac12\boldsymbol\beta\times(\boldsymbol\beta\times\boldsymbol\mu),
 \end{aligned}
 \label{eq:C-ED-MD-moments}
\end{equation}
with $\mathsf S=\mathsf Q=0$ and $\mathbf T=0$.  Inserted into
the terms written out in Eq.~\eqref{eq:B-eq8}, these moments reproduce the exact currents at every
$\mathbf k$.  For the ED, the difference between the two is
$\tfrac{ic}{2}[(\boldsymbol\beta\cdot\mathbf k)(\mathbf d-\mathsf B\mathbf d)
+\boldsymbol\beta\,\mathbf k\cdot(\mathsf B\mathbf d-\mathbf d)]$, which
vanishes because $\mathsf B\mathbf d-\mathbf d$ is parallel to
$\boldsymbol\beta$, and for the MD it vanishes identically.
The terms written out in Eq.~\eqref{eq:B-eq8} are therefore exact for the ED
and the MD of moving matter at any $\beta<1$.  For $\mathbf d\perp\mathbf v$ and
$\hat{\mathbf n}=\pm\hat{\mathbf v}$, Eq.~\eqref{eq:C-ED-MD} gives
$\Pi\widetilde{\mathbf J}=-i(\omega_{s0}\pm c\beta k)\mathbf d\pm ic\beta k\mathbf d
=-i\omega_{s0}\mathbf d$, so the forward and backward transverse currents are
equal, and the forward tilt of the received pattern comes from the prefactor
of Eq.~\eqref{eq:A-power}.

The MD $\mathbf m=-\tfrac12\boldsymbol\beta\times\mathbf d$ of the moving ED is
the sum of the moments of its two currents.  The polarization current $\partial_t\boldsymbol{\mathcal P}$ of the
translating dipole and the magnetization current $c\nabla\times\boldsymbol{\mathcal M}_R$, with
$\boldsymbol{\mathcal M}_R=(\mathbf d\times\boldsymbol\beta)\delta_ve^{-i\omega_{s0}t}$,
carry
\begin{equation}
 \mathbf m_{\mathcal P}=+\tfrac12\boldsymbol\beta\times\mathbf d,\qquad
 \mathbf m_{\mathcal M}=-\boldsymbol\beta\times\mathbf d.
 \label{eq:C-PM-moments}
\end{equation}
The magnetization current is the R\"ontgen current~\cite{Rontgen1888,LandauLifshitzECM}, which arises because the
charges of the dipole move with $\mathbf v$, and $\mathbf m_{\mathcal M}$ is
the magnetic moment that a Lorentz transformation assigns to a moving
ED~\cite{Hnizdo2012}.  The synthetic ED has no magnetization, so its MD is
$+\tfrac12\boldsymbol\beta\times\mathbf d$, and
$\mathbf J_{\rm mat}-\mathbf J_{\rm syn}=c\nabla\times\boldsymbol{\mathcal M}_R
+O(\beta^2)$.  The split refers to the representation of Eq.~\eqref{eq:C-ED-MD}.  A redistribution
$\boldsymbol{\mathcal P}\to\boldsymbol{\mathcal P}+\nabla\times\mathbf g$,
$\boldsymbol{\mathcal M}\to\boldsymbol{\mathcal M}-c^{-1}\partial_t\mathbf g$
leaves $\rho$ and $\mathbf J$, and hence $\mathbf m$ and the radiation,
unchanged.

\subsection{The TD in moving matter}

For the TD, Eq.~\eqref{eq:C-boost} with $\widetilde\rho'=0$ and
$\widetilde{\mathbf J}'=c\mathbf w(\mathbf k')$ gives
$\widetilde{\mathbf J}=(c/\gamma)\mathsf B^{-1}\mathbf w(\mathbf k')$ and
$\widetilde\rho=\boldsymbol\beta\cdot\mathbf w(\mathbf k')$, both quadratic in
$\mathbf k$, with the moments
\begin{equation}
 \begin{gathered}
 \mathbf p=ik_0\big[\beta^2\boldsymbol\tau-\boldsymbol\beta(\boldsymbol\beta\cdot\boldsymbol\tau)\big],\quad
 \mathbf m=-\frac{ik_0}{2}\Big(1+\frac{2}{\gamma^2}\Big)\boldsymbol\beta\times\boldsymbol\tau,\\
 \mathsf S=-2(\boldsymbol\beta\cdot\boldsymbol\tau)\mathsf B^2
 +(\mathsf B\boldsymbol\beta)(\mathsf B\boldsymbol\tau)^{\mathsf T}
 +(\mathsf B\boldsymbol\tau)(\mathsf B\boldsymbol\beta)^{\mathsf T},\\
 \mathbf T=\frac{4+\gamma^{-2}}{5}\boldsymbol\tau_\parallel
 +\frac{3+2\gamma^{-2}}{5\gamma}\boldsymbol\tau_{\rm t},
 \end{gathered}
 \label{eq:C-TD-moments}
\end{equation}
where $\boldsymbol\tau_\parallel=\hat{\mathbf v}(\hat{\mathbf v}\cdot\boldsymbol\tau)$
and $\boldsymbol\tau_{\rm t}=\boldsymbol\tau-\boldsymbol\tau_\parallel$.
Through $O(\beta)$, $\mathbf p=O(\beta^2)$,
$\mathbf m=-\tfrac{3ik_0}{2}\boldsymbol\beta\times\boldsymbol\tau$,
$\mathsf Q=3(\boldsymbol\beta\boldsymbol\tau^{\mathsf T}+\boldsymbol\tau\boldsymbol\beta^{\mathsf T})
-2(\boldsymbol\beta\cdot\boldsymbol\tau)I$, and $\mathbf T=\boldsymbol\tau$.
The terms written out in Eq.~\eqref{eq:B-eq8} reproduce the parts of degree
zero and one in $\mathbf k$ exactly, but of the part of degree two only the TD
term.  The difference is
the MQ and the symmetric traceless part of Eq.~\eqref{eq:B-second}, which
here equal the MQ
$\mathsf M_2=\tfrac{2}{3\gamma}[\boldsymbol\beta(\boldsymbol\beta\times\boldsymbol\tau)^{\mathsf T}
+(\boldsymbol\beta\times\boldsymbol\tau)\boldsymbol\beta^{\mathsf T}]$ and the
convection term carried by the EQ, both $O(\beta^2)$.  For
$\boldsymbol\tau\perp\mathbf v$, the exact transverse current minus
the terms written out in Eq.~\eqref{eq:B-eq8} is
\begin{equation}
 \Delta\mathbf J_\perp=\frac{c\beta^2k(\hat{\mathbf n})^2}{\gamma}
 \Big[\frac25-(\hat{\mathbf v}\cdot\hat{\mathbf n})^2\Big]\boldsymbol\tau_\perp,
 \label{eq:C-TD-remainder}
\end{equation}
and its maximum over directions,
$3ck_0^2|\boldsymbol\tau|\beta^2/[5\gamma^3(1-\beta)^2]$, lies forward and
equals 0.00730, 0.0957, and 0.691 in units of $ck_0^2|\boldsymbol\tau|$ at
$\beta=0.1$, 0.3, and 0.6.  Since the third charge moment vanishes,
Eq.~\eqref{eq:B-completed} removes this difference for any $\beta<1$, with a
numerical residual below $1.1\times10^{-15}$ in these units.

\subsection{Currents through \texorpdfstring{$O(\beta)$}{O(beta)}}

For the ED in moving matter, through $O(\beta)$,
$\widetilde{\mathbf J}=-i\omega_0\mathbf d-ic\boldsymbol\beta(\mathbf d\cdot\mathbf k)$
and, at the light cone,
\begin{equation}
 \Pi(\hat{\mathbf n})\widetilde{\mathbf J}\big(k(\hat{\mathbf n})\hat{\mathbf n}\big)
 =-i\omega_0\mathbf d_\perp-i\omega_0d_n\boldsymbol\beta_\perp+O(\beta^2).
 \label{eq:C-ED-current}
\end{equation}
In Eq.~\eqref{eq:B-eq8}, the ED term gives $-i\omega_0\mathbf d_\perp$, the MD
term $-\tfrac{i\omega_0}{2}(d_n\boldsymbol\beta_\perp-\beta_n\mathbf d_\perp)$,
and the convection term
$-\tfrac{i\omega_0}{2}(d_n\boldsymbol\beta_\perp+\beta_n\mathbf d_\perp)$.
These three terms are Eq.~\eqref{eq:material-ed}, and their sum is
Eq.~\eqref{eq:C-ED-current}.  The MD term is $O(\beta)$, so evaluating it at
$k(\hat{\mathbf n})$ instead of $\omega_{s0}/c$ changes it only at
$O(\beta^2)$, and the conventional relations and the Doppler-shifted
relations both miss the convection term.  For $\mathbf d\perp\mathbf v$ they
give the forward and backward currents $-i\omega_0\mathbf d(1\mp\beta/2)$.

For the MD in moving matter, the ED and MD terms reproduce
$\Pi\widetilde{\mathbf J}=ick(\hat{\mathbf n})\hat{\mathbf n}\times\boldsymbol\mu
-i\omega_0\Pi(\boldsymbol\beta\times\boldsymbol\mu)+O(\beta^2)$ once the MD
term is taken at $k(\hat{\mathbf n})$.  For the TD, the exact current at the light cone is, through $O(\beta)$,
$\Pi\widetilde{\mathbf J}=ck_0^2(\boldsymbol\tau_\perp+\tau_n\boldsymbol\beta_\perp)+O(\beta^2)$.
Of this, the MD and EQ terms supply
$\tfrac32ck_0^2(\tau_n\boldsymbol\beta_\perp-\beta_n\boldsymbol\tau_\perp)$ and
$-\tfrac12ck_0^2(\tau_n\boldsymbol\beta_\perp+\beta_n\boldsymbol\tau_\perp)$,
and the rest, $2ck_0^2\beta_n\boldsymbol\tau_\perp$, is the Doppler change
$c[k(\hat{\mathbf n})^2-k_0^2]\boldsymbol\tau_\perp$ of the TD term.

\subsection{Synthetic motion}

The synthetic sources of Eq.~\eqref{eq:A-synthetic} have the exact amplitudes
\begin{equation}
 \begin{aligned}
 \text{ED:}\quad&\widetilde\rho=-i\mathbf d\cdot\mathbf k,\quad
 \widetilde{\mathbf J}=-i(\omega_0+\mathbf v\cdot\mathbf k)\mathbf d,\\
 \text{MD:}\quad&\widetilde{\mathbf J}=ic\mathbf k\times\boldsymbol\mu,\qquad
 \text{TD:}\quad\widetilde{\mathbf J}=c\,\mathbf w(\mathbf k),
 \end{aligned}
 \label{eq:C-synthetic}
\end{equation}
where $\omega_0+\mathbf v\cdot\mathbf k$ comes from $\partial_t(\delta_ve^{-i\omega_0t})$.
The ED has $\mathbf p=\mathbf d$, $\mathbf m=+\tfrac12\boldsymbol\beta\times\mathbf d$,
and $\mathsf S=0$, the MD only $\mathbf m=\boldsymbol\mu$, and the TD only
$\mathbf T=\boldsymbol\tau$.  For the ED, the ED, MD, and convection terms of
Eq.~\eqref{eq:B-eq8} sum to $-i(\omega_0+\mathbf v\cdot\mathbf k)\mathbf d$, so
the terms written out in Eq.~\eqref{eq:B-eq8} are exact for all three sources at any $\beta$.  On the
light cone, $\omega_0+\mathbf v\cdot\mathbf k=ck$, and for $\beta_n<1$
\begin{equation}
 \text{ED: }-ick(\hat{\mathbf n})\mathbf d_\perp,\quad
 \text{MD: }ick(\hat{\mathbf n})\hat{\mathbf n}\times\boldsymbol\mu,\quad
 \text{TD: }ck(\hat{\mathbf n})^2\boldsymbol\tau_\perp.
 \label{eq:C-synthetic-light-cone}
\end{equation}
For $\beta_n>1$, the conjugate component gives the same expressions with
$k_-$ and $\mathbf d^*$, $\boldsymbol\mu^*$, $\boldsymbol\tau^*$.  Through
$O(\beta)$ the conventional relations miss the same convection term for the
synthetic ED as for the ED in moving matter, although the MD has the opposite
sign.  The synthetic MD and TD carry no ED,
so their convection term vanishes and the Doppler wavenumber alone accounts
for their difference from the conventional relations, to all orders in
$\beta$.  For a superluminal ED pattern with $\mathbf d\perp\mathbf v$, the radiation
in the forward direction comes from the conjugate component at $k_-=k_0/(\beta-1)$,
and that in the backward direction from the original component at $k_0/(1+\beta)$, so the
ratio of the transverse-current magnitudes is $(\beta+1)/(\beta-1)$.

Each panel of Fig.~\ref{fig:dipoles} uses laboratory-frame moments computed
exactly at the given $\beta$, without expansion in $\beta$, and the prefactor of
Eq.~\eqref{eq:A-power}.  Its solid curves are obtained from the terms written out in
Eq.~\eqref{eq:B-eq8}, which are exact for every source shown except the TD in
moving matter, whose solid curve is obtained from Eq.~\eqref{eq:B-completed}.  Every solid curve therefore coincides with the
exact pattern.  For a dipole perpendicular to $\mathbf v$, the forward-to-backward
ratio of the transverse current is 1 for all three dipoles in moving matter,
$\eta$ for the ED and the MD in synthetic motion, and $\eta^2$ for the TD in
synthetic motion, and the prefactor raises these to the power ratios
$\eta^4$, $\eta^6$, and $\eta^8$.  For the MD and the TD in moving matter this
follows from the factor $D^4$ of Appendix~\ref{app:power} and a rest-frame
pattern symmetric under $\hat{\mathbf n}'\to-\hat{\mathbf n}'$.  At
$\beta=0.3$ the three power ratios are 11.9, 41.0, and 141.5, whereas the
conventional relations give $\eta^4[(1-\beta/2)/(1+\beta/2)]^2=6.50$ for the
ED in moving matter.

\subsection{The anapole carried by matter}
\label{app:anapole-moments}

The rest-frame amplitudes of the anapole are
$c\widetilde\rho'=-ck_0(\boldsymbol\tau\cdot\mathbf k')$ and
$\widetilde{\mathbf J}'=-ck_0^2\boldsymbol\tau+c\mathbf w(\mathbf k')$.  On the
rest-frame light cone $|\mathbf k'|=k_0$ its four-current is
$-c(\boldsymbol\tau\cdot\mathbf k')(k_0,\mathbf k')$, proportional to the
four-wavevector, so its transverse part vanishes in every direction and the
anapole does not radiate~\cite{DevaneyWolf1973,Savinov2018}.  The boost maps
this into a laboratory four-current proportional to $(\omega/c,\mathbf k)$ on
the laboratory light cone, so the anapole carried by matter remains nonradiating for any $\beta<1$, a consequence of Lorentz covariance.

Through $O(\beta)$, its laboratory current and moments are
\begin{equation}
 \begin{gathered}
 \widetilde{\mathbf J}=-ck_0^2\boldsymbol\tau+c\mathbf w(\mathbf k)
 -ck_0\big[2(\boldsymbol\beta\cdot\mathbf k)\boldsymbol\tau
 -\mathbf k(\boldsymbol\beta\cdot\boldsymbol\tau)\big],\\
 \mathbf p=-ik_0\boldsymbol\tau,\quad\mathbf T=\boldsymbol\tau,\quad
 \mathbf m=-ik_0\boldsymbol\beta\times\boldsymbol\tau=\boldsymbol\beta\times\mathbf d,\\
 \mathsf Q=3(\boldsymbol\beta\boldsymbol\tau^{\mathsf T}+\boldsymbol\tau\boldsymbol\beta^{\mathsf T})
 -2(\boldsymbol\beta\cdot\boldsymbol\tau)I,
 \end{gathered}
 \label{eq:C-anapole-moments}
\end{equation}
where $\mathbf m$ is the sum of $+\tfrac{ik_0}{2}\boldsymbol\beta\times\boldsymbol\tau$
from the ED and $-\tfrac{3ik_0}{2}\boldsymbol\beta\times\boldsymbol\tau$ from
the TD.  These moments refer to a fixed laboratory time.

\subsection{Cancellation in moving matter}

At $\mathbf k=k(\hat{\mathbf n})\hat{\mathbf n}$, the terms of
Eq.~\eqref{eq:B-eq8} with the moments of Eq.~\eqref{eq:C-anapole-moments} are,
through $O(\beta)$,
\begin{equation}
 \begin{aligned}
 \boldsymbol{\mathcal A}_{ET}&=2ck_0^2\beta_n\boldsymbol\tau_\perp,\qquad
 \boldsymbol{\mathcal A}_M=-ck_0^2(\beta_n\boldsymbol\tau_\perp-\tau_n\boldsymbol\beta_\perp),\\
 \boldsymbol{\mathcal A}_Q&=\boldsymbol{\mathcal A}_v
 =-\tfrac12ck_0^2(\beta_n\boldsymbol\tau_\perp+\tau_n\boldsymbol\beta_\perp),
 \end{aligned}
 \label{eq:C-anapole-terms}
\end{equation}
where $\boldsymbol{\mathcal A}_{ET}$ combines the ED and TD terms and
Eq.~\eqref{eq:A-doppler} has been used.  They sum to zero through $O(\beta)$, which is
Eq.~\eqref{eq:anapole-terms} with the ED and TD terms combined.  The conventional relations, evaluated at
the carrier wavenumber, keep the cancellation of the ED and TD terms and predict
$\boldsymbol{\mathcal A}_M+\boldsymbol{\mathcal A}_Q$ where the exact current
vanishes.  Evaluated at the Doppler wavenumber, they predict
$-\boldsymbol{\mathcal A}_v$.  For $\boldsymbol\tau\perp\mathbf v$ and
observation at $\theta=60^\circ$ in the plane of $\mathbf v$ and
$\boldsymbol\tau$, projected on the polar unit vector, the four terms in
units of $ck_0^2\beta|\boldsymbol\tau|$ are 0.5, $-1$, 0.25, and 0.25, the
slopes at small $\beta$ of the curves in Fig.~\ref{fig:anapole}(c).

For independent $\mathbf d$ and $\boldsymbol\tau$, the terms can instead be
grouped by the dipole that generates them.  The ED term, the induced MD
$-\tfrac12\boldsymbol\beta\times\mathbf d$, and the convection term carried by
$\mathbf p=\mathbf d$ give
$\boldsymbol{\mathcal A}_d=-ick_0(\mathbf d_\perp+d_n\boldsymbol\beta_\perp)+O(\beta^2)$
[Eq.~\eqref{eq:C-ED-current}].  The TD term at the Doppler wavenumber, its
induced MD, and its EQ give
\begin{equation}
 \begin{aligned}
 \boldsymbol{\mathcal A}_\tau={}&ck_0^2(1+2\beta_n)\boldsymbol\tau_\perp
 +\tfrac32ck_0^2(\tau_n\boldsymbol\beta_\perp-\beta_n\boldsymbol\tau_\perp)\\
 &-\tfrac12ck_0^2(\tau_n\boldsymbol\beta_\perp+\beta_n\boldsymbol\tau_\perp)
 =ck_0^2(\boldsymbol\tau_\perp+\tau_n\boldsymbol\beta_\perp),
 \end{aligned}
 \label{eq:C-TD-group}
\end{equation}
through $O(\beta)$.  Their sum depends on the same combination as at rest,
\begin{equation}
 \Pi\widetilde{\mathbf J}=-ick_0\big[\mathbf a_\perp+a_n\boldsymbol\beta_\perp\big]
 +O(\beta^2),\qquad\mathbf a=\mathbf d+ik_0\boldsymbol\tau,
 \label{eq:C-grouping}
\end{equation}
so motion preserves their common angular dependence.  In synthetic motion,
Eq.~\eqref{eq:C-synthetic-light-cone} gives instead
$\Pi\widetilde{\mathbf J}=-ick(\hat{\mathbf n})[\mathbf d_\perp+ik(\hat{\mathbf n})\boldsymbol\tau_\perp]$,
exact for $\beta_n<1$, in which the ED and TD terms scale as
$k(\hat{\mathbf n})$ and $k(\hat{\mathbf n})^2$.

With the moments of Eqs.~\eqref{eq:C-ED-MD-moments} and
\eqref{eq:C-TD-moments}, the terms written out in Eq.~\eqref{eq:B-eq8} reproduce the ED
current exactly, so its difference from the vanishing anapole current is the TD
remainder of Eq.~\eqref{eq:C-TD-remainder}.  The maximum of its received
power is 0.231\,\% of the maximum of the conventional prediction at
$\beta=0.1$ and 5.14\,\% at $\beta=0.3$.  Equation~\eqref{eq:B-completed}
cancels it for any $\beta<1$, with a residual below
$10^{-15}ck_0^2|\boldsymbol\tau|$ at $\beta=0.1$, 0.3, and 0.6.
Over $10^{-3}\le\beta\le0.9$, the error of the conventional relations is linear in
$\beta$, that of the terms written out in Eq.~\eqref{eq:B-eq8} quadratic, and
Eq.~\eqref{eq:B-completed} is exact to machine precision for both the TD and the anapole carried by
matter.  The solid curve in Fig.~\ref{fig:anapole}(a) uses the completed decomposition, and
that in Fig.~\ref{fig:anapole}(b) is already exact with the terms written out in Eq.~\eqref{eq:B-eq8}.

\subsection{The synthetic anapole}

For the synthetic ED and TD with $\mathbf d=-ik_0\boldsymbol\tau$,
Eq.~\eqref{eq:C-synthetic-light-cone} gives exactly
\begin{equation}
 \Pi(\hat{\mathbf n})\widetilde{\mathbf J}=
 \begin{cases}
 ck(\hat{\mathbf n})\big[k(\hat{\mathbf n})-k_0\big]\boldsymbol\tau_\perp,&\beta_n<1,\\
 ck_-\big(k_-+k_0\big)\boldsymbol\tau_\perp^*,&\beta_n>1,
 \end{cases}
 \label{eq:C-synthetic-anapole}
\end{equation}
the second line belonging to the conjugate component.  Since
$k-k_0=k_0\beta_n/(1-\beta_n)$, the original component vanishes only where
$\beta_n=0$, and the conjugate factor $k_-(k_-+k_0)$ is positive.  Apart from
directions with $\boldsymbol\tau_\perp=0$, the synthetic anapole therefore
radiates in every direction except in the plane perpendicular to $\mathbf v$.
This plane lies in the region $\beta_n<1$ also for $\beta>1$, so the null
plane persists for superluminal patterns.  Through $O(\beta)$ the terms are
$\boldsymbol{\mathcal A}_{ET}+\tfrac12\boldsymbol{\mathcal A}_M+\boldsymbol{\mathcal A}_v
=ck_0^2\beta_n\boldsymbol\tau_\perp$ in the notation of
Eq.~\eqref{eq:C-anapole-terms}.  The TD pattern acquires no MD or EQ, and the
MD of the ED pattern, $+\tfrac12\boldsymbol\beta\times\mathbf d$, has the sign
opposite to that in moving matter, so the MD term is half that in moving
matter and the EQ term is absent
[Fig.~\ref{fig:anapole}(d)].

For $\boldsymbol\tau\perp\mathbf v$ and observation in the plane of
$\mathbf v$ and $\boldsymbol\tau$, where
$|\boldsymbol\tau_\perp|^2=|\boldsymbol\tau|^2\cos^2\theta$,
Eqs.~\eqref{eq:A-power} and \eqref{eq:C-synthetic-anapole} give
\begin{equation}
 \frac{dP_{\rm rec}}{d\Omega}=\frac{ck_0^6|\boldsymbol\tau|^2}{8\pi}\,
 \frac{\beta^2\cos^4\theta}{(1-\beta\cos\theta)^8}
 \label{eq:C-synthetic-power}
\end{equation}
for $\beta<1$.  In an arbitrary direction, with
$\boldsymbol\tau\parallel\hat{\mathbf x}$ and $\mathbf v\parallel\hat{\mathbf z}$,
the angular factor $\beta^2n_z^2(1-n_x^2)/(1-\beta n_z)^8$ cannot exceed
$\beta^2/(1-\beta)^8$, its value at $+\hat{\mathbf z}$, so the maximum lies
along $+\mathbf v$ for every $0<\beta<1$.  Integrating over directions gives, at low speed,
\begin{equation}
 P_{\rm rec}=\frac25\beta^2P_{\rm ED}+O(\beta^3),\qquad
 P_{\rm ED}=\frac{\omega_0^4|\mathbf d|^2}{3c^3},
 \label{eq:C-synthetic-total}
\end{equation}
where $P_{\rm ED}$ is the power of the ED alone at rest [Fig.~\ref{fig:anapole}(b)].  For the
anapole carried by matter, the exact power vanishes, whereas the conventional
relations predict $1.150\,\beta^2P_{\rm ED}$, their evaluation at the Doppler
wavenumber $0.150\,\beta^2P_{\rm ED}$, and the terms written out in Eq.~\eqref{eq:moving-multipole}
$0.097\,\beta^4P_{\rm ED}$ (low-speed coefficients, computed numerically).

\paragraph{Restoring the nonradiating condition.}
The residual $ck_0^2\beta_n\boldsymbol\tau_\perp$ of a synthetic anapole can
be canceled through $O(\beta)$ by adding to the pattern, while keeping
$\mathbf d=-ik_0\boldsymbol\tau$ and $\mathbf T=\boldsymbol\tau$, the moments
\begin{equation}
 \mathbf m_{\rm add}=-\frac{ik_0}{2}\boldsymbol\beta\times\boldsymbol\tau,\qquad
 \mathsf Q_{\rm add}=3(\boldsymbol\beta\boldsymbol\tau^{\mathsf T}
 +\boldsymbol\tau\boldsymbol\beta^{\mathsf T})-2(\boldsymbol\beta\cdot\boldsymbol\tau)I.
 \label{eq:C-add}
\end{equation}
Their terms in Eq.~\eqref{eq:B-eq8},
$\tfrac{ck_0k}{2}(\tau_n\boldsymbol\beta_\perp-\beta_n\boldsymbol\tau_\perp)$
and $-\tfrac{ck_0k}{2}(\tau_n\boldsymbol\beta_\perp+\beta_n\boldsymbol\tau_\perp)$,
sum to $-ck_0k\beta_n\boldsymbol\tau_\perp$, which cancels the residual
through $O(\beta)$.  The laboratory MD of the pattern becomes
$\tfrac12\boldsymbol\beta\times\mathbf d+\mathbf m_{\rm add}
=-ik_0\boldsymbol\beta\times\boldsymbol\tau$, and its EQ becomes
$\mathsf Q_{\rm add}$, so both agree with those of the anapole carried by
matter in Eq.~\eqref{eq:C-anapole-moments}.  The added moments belong to the
pattern itself and move with it, on the same support
$\omega=\omega_0+\mathbf v\cdot\mathbf k$.  A stationary source cannot cancel
the residual, because it radiates at $\omega_0$ in every direction, whereas
the pattern radiates at $ck(\hat{\mathbf n})$, and the two meet only in the
plane $\beta_n=0$.  Since the added moments are proportional to $\beta$, the
pattern must be designed for its velocity.  They induce further moments only
at $O(\beta^2)$, through the convective part
$-(\mathbf v\cdot\nabla)\boldsymbol{\mathcal P}$ of a polarization current, and
none if the MD is realized as a magnetization pattern.

For $\beta<1$, the exact condition follows from the full four-current.  The
pattern's support $\omega=\omega_0+\mathbf v\cdot\mathbf k$ is that of a
source with rest frequency $\gamma\omega_0$ moving at $\mathbf v$, since the
boost to the frame moving at $\mathbf v$ gives
$\omega'=\gamma(\omega-\mathbf v\cdot\mathbf k)=\gamma\omega_0$.  Continuity
makes a nonradiating four-current proportional to the four-wavevector on the
light cone, a property preserved by the boost.  The pattern is therefore
nonradiating if and only if its laboratory current is that of a nonradiating
source at $\gamma k_0$ in uniform motion.  This condition is imposed on the
total laboratory current and includes all induced moments.  For a point
anapole it requires the complete second current moment, as in
Eq.~\eqref{eq:B-completed}.  For $\beta>1$ no such source exists, and the
condition must be imposed in the laboratory frame on both branches of
Eq.~\eqref{eq:C-synthetic-anapole}.

\section{Nonuniform translation}
\label{app:nonuniform}

\subsection{Moments about the instantaneous center}

Let the center of the source follow a twice-differentiable trajectory
$\mathbf R(t)$ with velocity $\mathbf v(t)=\dot{\mathbf R}(t)$.  The moments of
Eq.~\eqref{eq:A-moments} are now taken in the coordinates
$\mathbf r=\mathbf x-\mathbf R(t)$ and depend on time, and
$\mathbf I(t)=\int\mathbf J\,d^3x$ is the net current.  At fixed laboratory
position $\mathbf x$, $\dot r_a=-v_a$.  Multiplying
$\partial_t\rho+\nabla\cdot\mathbf J=0$ by 1, $r_a$, and $r_ar_b$ and
integrating, with no boundary terms for a localized source and harmonic or
periodic sources understood as distributions, gives exactly
\begin{equation}
 \dot q=0,\qquad \mathbf I=\dot{\mathbf p}+q\mathbf v,\qquad
 M_{ab}+M_{ba}=\dot S_{ab}+v_ap_b+v_bp_a,
 \label{eq:D-moments}
\end{equation}
for any speed and acceleration~\cite{Ribaric1995}.  For constant $\mathbf v$ and moments
proportional to $e^{-i\omega_{s0}t}$, the last relation reduces to
Eq.~\eqref{eq:B-translating}.  The same result follows from the moments about
a fixed origin by the shifts $\mathbf p=\mathbf p^{(0)}-q\mathbf R$ and their
analogues.

\subsection{The radiated amplitude}

With the center current
$\mathbf j_c(\mathbf k,t)=\int d^3r\,e^{-i\mathbf k\cdot\mathbf r}\mathbf J(\mathbf R+\mathbf r,t)$
and the trajectory phase $\mathcal E=e^{i\omega t-i\mathbf k\cdot\mathbf R(t)}$,
the space--time amplitude is
$\widetilde{\mathbf J}(\omega,\mathbf k)=\int dt\,\mathcal E\,\mathbf j_c$, and
the field radiated in direction $\hat{\mathbf n}$ at frequency $\omega>0$
follows from Eq.~\eqref{eq:A-farfield} with the transverse current
$\Pi(\hat{\mathbf n})\widetilde{\mathbf J}(\omega,(\omega/c)\hat{\mathbf n})$.
Expanding $e^{-i\mathbf k\cdot\mathbf r}$ and using Eq.~\eqref{eq:D-moments}
gives
\begin{equation}
 \mathbf j_c=\dot{\mathbf p}+q\mathbf v+ic\mathbf k\times\mathbf m
 -\frac i2\dot{\mathsf S}\mathbf k
 -\frac i2\big[\mathbf v(\mathbf p\cdot\mathbf k)+\mathbf p(\mathbf v\cdot\mathbf k)\big]
 +\boldsymbol{\mathcal R}_2,
 \label{eq:D-jc}
\end{equation}
where the remainder $\boldsymbol{\mathcal R}_2$ of degree two and higher obeys
$|\boldsymbol{\mathcal R}_2|\le\tfrac12|\mathbf k|^2\int r^2|\mathbf J|\,d^3r$
for an integrable current.  The convection term keeps its form, and the time
derivative takes the place of $-i\omega_{s0}$.  Integrating by parts,
$\int dt\,\mathcal E\dot f=-i\int dt\,\mathcal E[\omega-\mathbf k\cdot\mathbf v(t)]f$,
so that
\begin{equation}
 \begin{aligned}
 \widetilde{\mathbf J}(\omega,\mathbf k)=\int\! dt\,\mathcal E\Big\{&
 -i(\omega-\mathbf k\cdot\mathbf v)\mathbf p+q\mathbf v+ic\mathbf k\times\mathbf m\\
 &-\tfrac12(\omega-\mathbf k\cdot\mathbf v)\mathsf S\mathbf k
 +\boldsymbol{\mathcal R}_2\\
 &-\tfrac i2\big[\mathbf v(\mathbf p\cdot\mathbf k)+\mathbf p(\mathbf v\cdot\mathbf k)\big]
 \Big\}.
 \end{aligned}
 \label{eq:D-amplitude}
\end{equation}
The trajectory phase is kept whole, and only the internal size of the source
about its center enters the expansion, so the excursion of the center may be
large compared with the wavelength.

\subsection{Instantaneous support and neutral patterns}

For a harmonic pattern, the phase $\Phi(t)=(\omega-\omega_0)t-\mathbf k\cdot\mathbf R(t)$
is stationary where
\begin{equation}
 \omega=\omega_0+\mathbf k\cdot\mathbf v(t),
 \label{eq:D-stationary}
\end{equation}
the instantaneous support of Sec.~\ref{sec:scope}.  If $\Delta\mathbf R$ is
the excursion over the time on which the velocity changes, a stationary-phase
interpretation requires a large projected phase excursion,
$|\mathbf k\cdot\Delta\mathbf R|\gg1$.  For sinusoidal translation this
parameter is $k_0\xi_0|n_x|$ near the carrier, and it fails for directions
with $n_x=0$.  Degenerate stationary points, such as turning points of the
instantaneous Doppler shift, require the full phase integral.

The rigid patterns considered here are localized polarization and
magnetization, with $\rho=-\nabla\cdot\boldsymbol{\mathcal P}$ and
$\mathbf J=\partial_t\boldsymbol{\mathcal P}+c\nabla\times\boldsymbol{\mathcal M}$,
and hence $q=0$.  There is then no acceleration term tied to a transported net
charge, and once the laboratory polarization and magnetization are prescribed
along the trajectory, acceleration enters the amplitude only through the
trajectory phase.  This does not prescribe the four-current of accelerated
matter.  For a prescribed laboratory ED
$\mathbf d(t)$ with $\delta_R=\delta^3(\mathbf x-\mathbf R(t))$, the synthetic
polarization $\boldsymbol{\mathcal P}=\mathbf d\,\delta_R$ gives
\begin{equation}
 \begin{gathered}
 \mathbf J_{\rm syn}=\dot{\mathbf d}\,\delta_R-\mathbf d(\mathbf v\cdot\nabla\delta_R),
 \\
 \widetilde{\mathbf J}_{\rm syn}(\omega,\mathbf k)=-i\omega\int dt\,\mathcal E\,\mathbf d(t),
 \end{gathered}
 \label{eq:D-synthetic-ED}
\end{equation}
Its moments are $\mathbf p=\mathbf d$, $M_{ab}=d_av_b$, and
$\mathbf m=\mathbf v\times\mathbf d/(2c)$.  The transform follows because the
MD and convection terms sum to $-i(\mathbf k\cdot\mathbf v)\mathbf d$ and the
integration by parts turns $\dot{\mathbf d}$ into
$-i(\omega-\mathbf k\cdot\mathbf v)\mathbf d$.  The current
$\dot{\mathbf d}\,\delta_R-\mathbf v(\mathbf d\cdot\nabla\delta_R)$ has the same
charge and the same symmetric moment but $\mathbf m=-\mathbf v\times\mathbf d/(2c)$,
and it reproduces the ED of moving matter through $O(\beta)$
[Eq.~\eqref{eq:C-ED-MD-moments}], whereas matter accelerated to high speed
requires the four-current fixed by its motion.

\subsection{Synthetic anapole on a trajectory}

The synthetic anapole on the trajectory $\mathbf R(t)$ is
$\boldsymbol{\mathcal P}=\mathbf d(t)\delta_R$ and
$\boldsymbol{\mathcal M}=\nabla\times[\boldsymbol\tau(t)\delta_R]$ with
$\mathbf d(t)=\dot{\boldsymbol\tau}(t)/c$, which for
$\boldsymbol\tau=\boldsymbol\tau_0e^{-i\omega_0t}$ gives
$\mathbf d=-ik_0\boldsymbol\tau$.  Its ED part,
$-(\omega/c)\int dt\,\mathcal E(\omega-\mathbf k\cdot\mathbf v)\boldsymbol\tau$
after integration by parts, and its TD part,
$c|\mathbf k|^2\Pi\int dt\,\mathcal E\boldsymbol\tau$, sum at
$\mathbf k=(\omega/c)\hat{\mathbf n}$ to, for any trajectory and speed,
\begin{equation}
 \boldsymbol{\mathcal A}(\omega,\hat{\mathbf n})=\frac{\omega^2}{c^2}\Pi(\hat{\mathbf n})
 \int dt\,e^{i\omega[t-\hat{\mathbf n}\cdot\mathbf R(t)/c]}
 \big(\hat{\mathbf n}\cdot\mathbf v(t)\big)\boldsymbol\tau(t).
 \label{eq:D-anapole-trajectory}
\end{equation}
For one harmonic component, with
$G(\omega,\mathbf k)=\int dt\,e^{i(\omega-\omega_0)t-i\mathbf k\cdot\mathbf R(t)}$,
\begin{equation}
 \boldsymbol{\mathcal A}=\frac{\omega(\omega-\omega_0)}{c}\,
 \boldsymbol\tau_{0\perp}\,G\Big(\omega,\frac\omega c\hat{\mathbf n}\Big),
 \label{eq:D-anapole-G}
\end{equation}
where the factor $\omega-\omega_0$ expresses the cancellation of the ED and TD
at the carrier frequency.  For $\mathbf R=\mathbf vt$,
$G=2\pi\delta(\omega-\omega_0-\mathbf k\cdot\mathbf v)$, and
Eq.~\eqref{eq:C-synthetic-anapole} follows.

\subsection{Periodic translation}

For $\mathbf R(t)=\xi_0\sin(\Omega t)\hat{\mathbf x}$ with $\xi_0\ge0$,
$\Omega>0$, and $n_x=\hat{\mathbf n}\cdot\hat{\mathbf x}$, the expansion
$e^{-iz\sin\Omega t}=\sum_\ell J_\ell(z)e^{-i\ell\Omega t}$ gives
$G=2\pi\sum_\ell J_\ell(k_x\xi_0)\delta(\omega-\omega_\ell)$ with
$\omega_\ell=\omega_0+\ell\Omega$.  Each line is evaluated at
$\mathbf k=(\omega_\ell/c)\hat{\mathbf n}$.  For the ED pattern
$\mathbf d(t)=\mathbf d_0e^{-i\omega_0t}$, Eq.~\eqref{eq:D-synthetic-ED} gives
exactly, with $z_\ell=\omega_\ell\xi_0n_x/c$,
\begin{equation}
 \boldsymbol{\mathcal A}_{\rm ED}=2\pi\!\!\sum_{\ell:\,\omega_\ell>0}\!\!
 \mathbf a_{\ell}\,\delta(\omega-\omega_\ell),\qquad
 \mathbf a_\ell=-i\omega_\ell J_\ell(z_\ell)\mathbf d_{0\perp},
 \label{eq:D-ED-lines}
\end{equation}
which is Eq.~\eqref{eq:sidebands}.  The ED term contributes
$-i\omega_0J_\ell\mathbf d_{0\perp}$.  At each time
$\mathbf m=\mathbf v(t)\times\mathbf d/(2c)$, so the MD and convection terms
together are $-i(\mathbf k\cdot\mathbf v)\mathbf d_\perp$, and the identity
$(\mathbf k\cdot\mathbf v)e^{-i\mathbf k\cdot\mathbf R}=i\partial_te^{-i\mathbf k\cdot\mathbf R}$
turns their contribution to line $\ell$ into $-i\ell\Omega J_\ell\mathbf d_{0\perp}$.
No expansion in $\xi_0$, $\Omega$, or the observation angle was made.  For
$n_x=0$, $J_\ell(0)=\delta_{\ell0}$ and only the carrier remains.  The result
holds also for $\xi_0\Omega\ge c$.

For the synthetic anapole, Eq.~\eqref{eq:D-anapole-G} gives instead
\begin{equation}
 \mathbf a_\ell=\frac{\omega_\ell\,\ell\Omega}{c}\,
 J_\ell(z_\ell)\,\boldsymbol\tau_{0\perp}.
 \label{eq:D-anapole-lines}
\end{equation}
The carrier coefficient $\mathbf a_0$ vanishes, because there the ED and TD
still cancel, whereas the sidebands are generally nonzero.  They vanish for
$\xi_0\to0$, for $n_x=0$ or $\boldsymbol\tau_{0\perp}=0$, and at zeros of
$J_\ell$.  Compared with the ED alone on the same trajectory, each line carries the extra amplitude factor $\ell\Omega/\omega_0$.  A
real source also contains the conjugate component, whose sidebands fall at
$-\omega_0+\ell\Omega$.  When $2\omega_0/\Omega$ is an integer, one of them
falls at $\omega_0$ and adds to the carrier.  Acceleration enters through the
trajectory phase, although Eq.~\eqref{eq:D-moments} contains no acceleration.
Accelerated anapoles carried by matter can also radiate~\cite{Savinov2018}, with a
four-current specified by the motion of the matter.

\section{Finite size}
\label{app:finite}

\subsection{Subluminal sources}

For a source of size $a$, the expansion about $\mathbf k=0$ requires
$k(\hat{\mathbf n})a\ll1$, with the remainder bounded as in
Eq.~\eqref{eq:D-jc}.  For a synthetic ED perpendicular to $\mathbf v$, the
point-source transverse currents have the forward-to-backward ratio $\eta$ of
Eq.~\eqref{eq:A-eta}, while the forward wavenumber $k_0/(1-\beta)$ grows with
$\beta$.  Requiring $k(\hat{\mathbf v})a\le1$ gives $1-\beta\ge k_0a$ and
\begin{equation}
 \eta\le\frac{2-k_0a}{k_0a}\simeq\frac{2}{k_0a}.
 \label{eq:E-bound}
\end{equation}
The bound is on the current-amplitude ratio $\eta$, and the ratios in Sec.~\ref{sec:dipoles} that contain $\eta$ must be read with this restriction for a
finite pattern.  For moving matter the forward wavenumber is instead
$k_0/[\gamma(1-\beta)]=k_0\sqrt{(1+\beta)/(1-\beta)}$, and for a specified
laboratory size the threshold $ka=1$ is at
\begin{equation}
 \beta=\frac{1-(k_0a)^2}{1+(k_0a)^2},\qquad
 1-\beta=\frac{2(k_0a)^2}{1+(k_0a)^2}\simeq2(k_0a)^2 .
 \label{eq:E-matter}
\end{equation}
The estimate $1-\beta\simeq k_0a$ thus applies to synthetic motion.  If the
size is specified in the rest frame, the direction-dependent laboratory size
must be transformed first.  For a pattern with the profile $f(\mathbf r)$, the
current is the point current multiplied by the form factor
$\widetilde f(\mathbf k)$, which for a Gaussian is
$F=e^{-|\mathbf k|^2a^2/2}$.  At $\beta=1-k_0a$ this reduces the point-source
ratios 39.0, 19.0, and 9.0 for $k_0a=0.05$, 0.1, and 0.2 to 23.7, 11.5, and
5.49, close to the largest ratios of the Gaussian source over all $\beta$.
Beyond $\beta=1-k_0a$
the ratio of the finite pattern falls, because the forward Doppler wavenumber
exceeds $1/a$ and the low-order moments no longer fix the radiation.

\subsection{Superluminal patterns}

\begin{figure}
 \includegraphics[width=0.85\columnwidth]{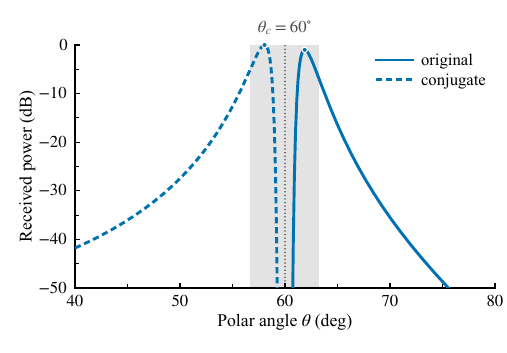}
 \caption{Received power of a Gaussian synthetic ED perpendicular to
 $\mathbf v$, with $\beta=2$ and $k_0a=0.1$, in the plane of $\mathbf v$ and
 the dipole, in decibels relative to its maximum, versus the polar angle
 $\theta$. The vertical line marks the Cherenkov direction $\theta_c=60^\circ$,
 and the shaded band the directions with $ka>1$, Eq.~\eqref{eq:E-band}. The
 conjugate component (dashed) radiates at $\theta<\theta_c$ and the original
 component (solid) at $\theta>\theta_c$. Their peaks lie at $58.05^\circ$ and
 $61.87^\circ$.}
 \label{fig:cherenkov}
\end{figure}

For $\beta>1$, $k=k_0/|1-\beta_n|$ diverges at the Cherenkov directions.  The
band $ka>1$ is $|1-\beta\cos\theta|<k_0a$, with edges
\begin{equation}
 \theta_\mp=\arccos\frac{1\pm k_0a}{\beta},\qquad
 \theta_+-\theta_-=\frac{2k_0a}{\beta\sin\theta_c}=\frac{2k_0a}{\sqrt{\beta^2-1}},
 \label{eq:E-band}
\end{equation}
where the width holds up to $O\big((k_0a)^3\big)$ and requires
$k_0a\ll\beta-1$.  For a Gaussian synthetic ED,
$|\Pi\widetilde{\mathbf J}|\propto k|F(ka)\mathbf d_\perp|$, and the prefactor
of Eq.~\eqref{eq:A-power} adds $k^4$, so
\begin{equation}
 \frac{dP_{\rm rec}}{d\Omega}\propto(ka)^6e^{-(ka)^2}|\mathbf d_\perp|^2,
 \qquad ka=\frac{k_0a}{|1-\beta_n|},
 \label{eq:E-band-power}
\end{equation}
whose radial factor peaks at $ka=\sqrt3$, near but not at the Cherenkov
directions.  For $\beta=2$ and $k_0a=0.1$, in the plane of $\mathbf v$ and the
dipole, the peaks of the original and conjugate components lie at
$61.87^\circ$ and $58.05^\circ$, on either side of $\theta_c=60^\circ$
(Fig.~\ref{fig:cherenkov}).  For $\beta=1.5$, 2, and 3 and $k_0a=0.05$, 0.1,
and 0.2, the band $ka>1$ carries 84.8\,\% to 84.9\,\% of the total power.
Setting $F=1$ overestimates the power by $e^{(ka)^2}$, a relative error of
1.72 at $ka=1$, so a low-order multipole decomposition is quantitative only for $ka\ll1$.

\subsection{A finite anapole}

Convolving the complete point anapole with a normalized nonnegative radial
profile $f(r)$ multiplies its four-current by $\widetilde f(\mathbf k)$, which
keeps the four-current of Sec.~\ref{app:anapole-moments} proportional to the
four-wavevector, and hence keeps the anapole nonradiating at rest for any size.  With
$\langle r^2\rangle=\int r^2f\,d^3r$,
$\widetilde f=1-\langle r^2\rangle|\mathbf k|^2/6+O(|\mathbf k|^4)$, and the
extra quadratic current $\tfrac c6k_0^2\langle r^2\rangle|\mathbf k|^2\boldsymbol\tau$
adds $\tfrac16k_0^2\langle r^2\rangle\boldsymbol\tau$ to $\mathbf T$ through
Eq.~\eqref{eq:A-fourier-moments}, so that
\begin{equation}
 \mathbf p=-ik_0\boldsymbol\tau,\qquad
 \mathbf p+ik_0\mathbf T=\frac{ik_0^3\langle r^2\rangle}{6}\boldsymbol\tau.
 \label{eq:E-finite-anapole}
\end{equation}
The two-moment condition of Eq.~\eqref{eq:A-anapole} thus holds up to the
relative correction $k_0^2\langle r^2\rangle/6$, which equals $(k_0a)^2/10$
for a uniform ball of radius $a$ and lies between 0 and $(k_0a)^2/6$ for any
profile supported within $r\le a$, and the higher moments restore the exact
zero.  This correction is present at $\beta=0$ and vanishes as $a\to0$,
whereas the motional terms of Appendix~\ref{app:dipoles} are present at $a=0$
and vanish as $\beta\to0$.

\bibliography{references}

\end{document}